\documentclass[a4paper,11pt]{article}

\usepackage{mystyle}
\usepackage{float}
\graphicspath{{images/}}

\newcommand{\D}[1]{(\mathbf{1}_{#1},0)}
\global\long\def\smb#1{\mathbf{S}_{#1}^{mb}}%
\global\long\def\sam#1{\mathbf{S}_{#1}^{am}}%
\global\long\def\scb#1{\mathbf{S}_{#1}^{cb}}%
\global\long\def\sck#1{\mathbf{S}_{#1}^{ck}}%
\global\long\def\skb#1{\mathbf{S}_{#1}^{kb}}%
\global\long\def\sab#1{\mathbf{S}_{#1}^{ab}}%
\global\long\def\sba#1{\mathbf{S}_{#1}^{ba}}%
\global\long\def\ska#1{\mathbf{S}_{#1}^{ka}}%
\global\long\def\sbk#1{\mathbf{S}_{#1}^{bk}}%
\global\long\def\tk#1{\mathbf{T}_{#1}^{k}}%
\global\long\def\ta#1{\mathbf{T}_{#1}^{a}}%
\global\long\def\tb#1{\mathbf{T}_{#1}^{b}}%
\global\long\def\tc#1{\mathbf{T}_{#1}^{c}}%
\global\long\def\td#1{\mathbf{T}_{#1}^{d}}%
\global\long\def\tn#1{\mathbf{T}_{#1}^{n}}%

\vspace{1cm}

\begin{document}
	\begin{titlepage}
			\hypersetup{allcolors=indigo}
		\noindent Preprint \hfill \today
		\vspace*{\fill}
		\begin{flushleft}
			{\Large \bf Colour structure of next-to-eikonal correlator webs at three loops  } \bigskip \bigskip  \\
			\bfseries\sffamily {Abhinava Danish$\,^a$, 
				Shubham Mishra$\,^{a, b}$, 
				Sourav Pal$\,^c$, 
				Aditya Srivastav$\,^a$,
				Anurag Tripathi$\,^a$ }
		\end{flushleft}
		\vspace{0.5cm} 
		{\it
			$\,^a$Department of Physics, Indian Institute of Technology Hyderabad, \\ Kandi, Sangareddy, Telangana State 502284, India   \medskip \\
			$\,^b$Department of Physics, School of Basic and Applied Science, 
			Galgotias University, Greater Noida, Uttar Pradesh 203201, India  \medskip \\
			$\,^c$School of Physical Sciences, National Institute of Science Education and Research, \\
			An OCC of Homi Bhaba National Institute, Jatni 752050, India  \medskip \\
		}
		\vspace{0.5cm}
		
		\noindent {\it E-mail}: \href{mailto:abhinavadan@gmail.com}{abhinavadan@gmail.com},
		\href{mailto:shubhammishra@galgotiasuniversity.edu.in}{shubhammishra@galgotiasuniversity.edu.in},
		\href{mailto:sourav@niser.ac.in}{sourav@niser.ac.in}, \\ \href{mailto:shrivastavadi333@gmail.com}{shrivastavadi333@gmail.com},  \href{mailto:tripathi@phy.iith.ac.in}{tripathi@phy.iith.ac.in}
		\bigskip \\
		\noindent {Abstract}: 
	     Correlators of Wilson lines, which capture eikonal contributions, are known to exponentiate in non-abelian gauge theories, and their logarithms can be organized in terms of collections of Feynman diagrams called webs. In~\cite{Agarwal:2020nyc}, the concept of a correlator web (Cweb) -- a set of skeleton diagrams built from connected gluon correlators, generalizing the notion of webs -- was introduced. The part of the next-to-eikonal contributions to the scattering amplitude that exponentiates is given in terms of next-to-eikonal Cwebs~\cite{Gardi:2010rn,Laenen:2008gt}.
	     In the present article, we study next-to-eikonal Cwebs at three-loop order, the order at which non-trivial Cweb mixing matrices first appear. The methods developed in~\cite{Agarwal:2022wyk} to construct mixing matrices directly, without the use of the replica trick, are used in this article to obtain the mixing matrices for next-to-eikonal Cwebs after we establish a relationship between next-to-eikonal and eikonal Cwebs.
		\vspace*{\fill}
	\end{titlepage}

	\hrule
	\hypersetup{allcolors=black}
	\tableofcontents
	\vspace{0.4cm}
	\hrule 	
	
	\hypersetup{allcolors=indigo}
	
	\section{Introduction}
	\label{sec:intro}
	Scattering amplitudes involving massless gauge bosons suffer from infrared singularities. These singularities can be factorised from the hard interaction part. As a consequence of factorisation, the scattering amplitude in the soft limit exponentiates as~\cite{Gatheral,Gatheral:1983cz,Gardi:2010rn,Laenen:2008gt}, 
	\begin{align}
		\mathcal{M}\,=\, \mathcal{M}_{0} \exp{[W]}\,=\,{\cal M}_0\exp\left[\sum_{d_{\text{E}},d'_{\text{E}}}
		K(d_{\text{E}}) \,R^{\text{E}}_{dd'}\, C(d'_{\text{E}}) \right] \,,
		\label{eq:eikW}
	\end{align} 
	where $ W $ is known as Cwebs (webs). A Cweb is made out of a set of diagrams ($ d $) that is closed under permutation of attachments on each Wilson line. The diagrams mix among themselves via Cweb mixing matrices $ R $.  The subscript $ E $ indicates that all the diagrams in these Cwebs are constructed using eikonal (soft) Feynman rules. The mixing matrices for eikonal Cwebs have three general properties: 
	\begin{itemize}
		\item [(i)] They are idempotent, that is, $ R^2=R $. 
		\item [(ii)] The entries of $ R $ obey the zero row sum rule, $ \sum_{d'} R(d,d')\,=\,0 $. 
		\item[(iii)] They also obey a conjectured column-sum rule  
		$\sum_{d}s(d)R(d,d')\,=\,0 $, where symmetry factor $ s(d) $ for a diagram $ d $ is defined as the number of
		ways in which the gluon correlators can be sequentially shrunk to their common origin. Symmetry factors for all the diagrams of a Cweb form a set called column weight vector $ S $.  
	\end{itemize}
	For a detailed discussion on these properties and their physical implications, see refs~\cite{Agarwal:2020nyc,Agarwal:2022wyk,Agarwal:2022xec,Gardi:2010rn,Gardi:2011wa,Mishra:2023acr,White:2015wha}.
	
	In recent years, there has been a growing interest in studying scattering amplitudes and cross-sections beyond the eikonal (soft) approximation. These corrections are formally known as next-to-eikonal  corrections as they are one order suppressed in powers of soft gluon's momentum,	
	as compared to the eikonal approximation. In the context of physical observables, such as cross-section, it has been shown that these corrections have sizeable numerical contributions~\cite{Kramer:1996iq,Ball:2013bra,Bonvini:2014qga,Anastasiou:2015ema,Anastasiou:2016cez,vanBeekveld:2019cks,vanBeekveld:2021hhv,Ajjath:2021lvg,Das:2024pac,Das:2025wbj,Bhattacharya:2025rqk}, and therefore several methods have been developed to calculate them~\cite{DelDuca:2017twk,Pal:2023vec,Pal:2024eyr,vanBeekveld:2019prq}. In presence of next-to-eikonal  corrections, the form of eq.~\eqref{eq:eikW}  changes to~\cite{Laenen:2008gt,Gardi:2010rn}, 
	\begin{equation}\label{eq:Fact-M-NE1}
		{\cal M}={\cal M}_0\exp\left[\sum_{d_{\text{E}},d'_{\text{E}}}
		K(d_{\text{E}}) \,R^{\text{E}}_{dd'}\, C(d'_{\text{E}})
		\,\,+
		\sum_{d_{\text{NE}},d'_{\text{NE}}}K(d_{\text{NE}})\,R^{\text{NE}}_{dd'}\,C(d'_{\text{NE}})\,
		\right](1+{\cal M}_r)\,.
	\end{equation}
	Here the second term in the square bracket is formally known as next-to-eikonal (NE) Cwebs. Here $ M_r $ denotes a remainder term that does not exponentiate. This term was originally introduced by Low in~\cite{Low:1958sn} and was generalised in~\cite{Burnett:1967km,DelDuca:1990gz}. Very recently, NE Cwebs are generalised for soft quark corrections in~\cite{vanBeekveld:2023liw}, however, in this article, we only consider next-to-soft gluon contributions. The quantity $ \sum_{d'}^{}R_{dd'}\,C(d') $ is referred as exponentiated colour factor (ECF) for the diagram $ d $ as this is the colour structure which sits in the exponent. 
	
	The central quantity in eq.~\eqref{eq:Fact-M-NE1} is the Cweb mixing matrix $ R$, that can be calculated using a replica trick algorithm for both the eikonal and NE Cwebs, as  shown in~\cite{Gardi:2010rn}. A method to calculate them without using replica trick is developed in~\cite{Agarwal:2022wyk}. In this article, we explore how the methods developed in~\cite{Agarwal:2022wyk} can be used to predict the mixing matrices for NE Cwebs as well. 
	
	The rest of this article is structured as follows. In section~\ref{sec:EandNE}, we relate eikonal and NE Cwebs purely using the objects of graph theory, and present an algorithm on how NE Cwebs can be generated at a given perturbative order. In section~\ref{sec:DNE}, we present the notation for NE Cwebs and how the mixing matrices for NE Cwebs can be directly computed using the results and concepts developed for Cwebs in~\cite{Agarwal:2022wyk}. Finally, in section~\ref{sec:results}, we present our summary and future outlook.    

\section{Eikonal and NE Cwebs: A combinatorial view}\label{sec:EandNE}

In this section, we establish two useful correspondences between eikonal and NE Cwebs,  which we will employ 
 to implement the formalism developed for eikonal Cwebs for the analysis of NE Cwebs. The first correspondence,
  \textit{identicality of shuffles} 
establishes a relation between the shuffle of gluon attachments on the Wilson lines for the eikonal and NE Cwebs. The second correspondence \textit{identicality of posets} connects the posets defined for eikonal Cwebs to those of NE Cwebs, leading to the definition of the symmetry factor ($ s $-factor) for NE Cweb diagrams.

\subsection{Identicality of shuffles}\label{sec:EandNEa}

The computation of mixing matrices for NE Cwebs is not fundamentally different from that of the eikonal Cwebs. The structure of Cweb mixing matrices depend only on the shuffle of the attachments to Wilson lines, and does not depend on the vertices away from the Wilson lines. The only difference between NE Cwebs and eikonal Cweb  is that the NE Cwebs are made out of seagull vertices instead of QCD vertices; in appendix~\eqref{sec:NEfeynmanrules} we give details of the two NE vertices --- single gluon vertex and seagull vertex.

\begin{figure}[H]
	\centering
	\subfloat[][]{\includegraphics[scale=0.06]{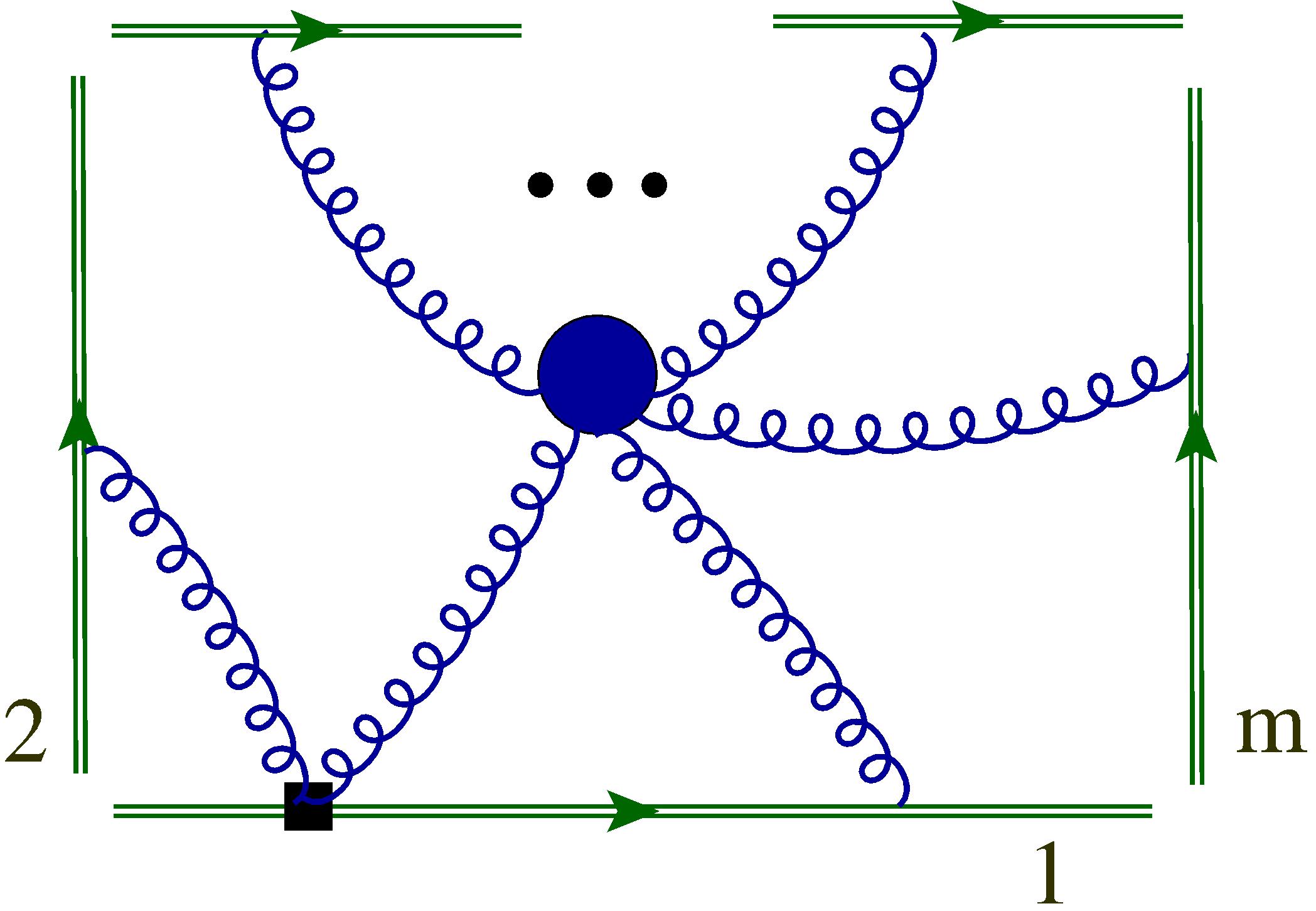} }
	\qquad \hspace{1cm}
	\subfloat[][ ]{\includegraphics[scale=0.06]{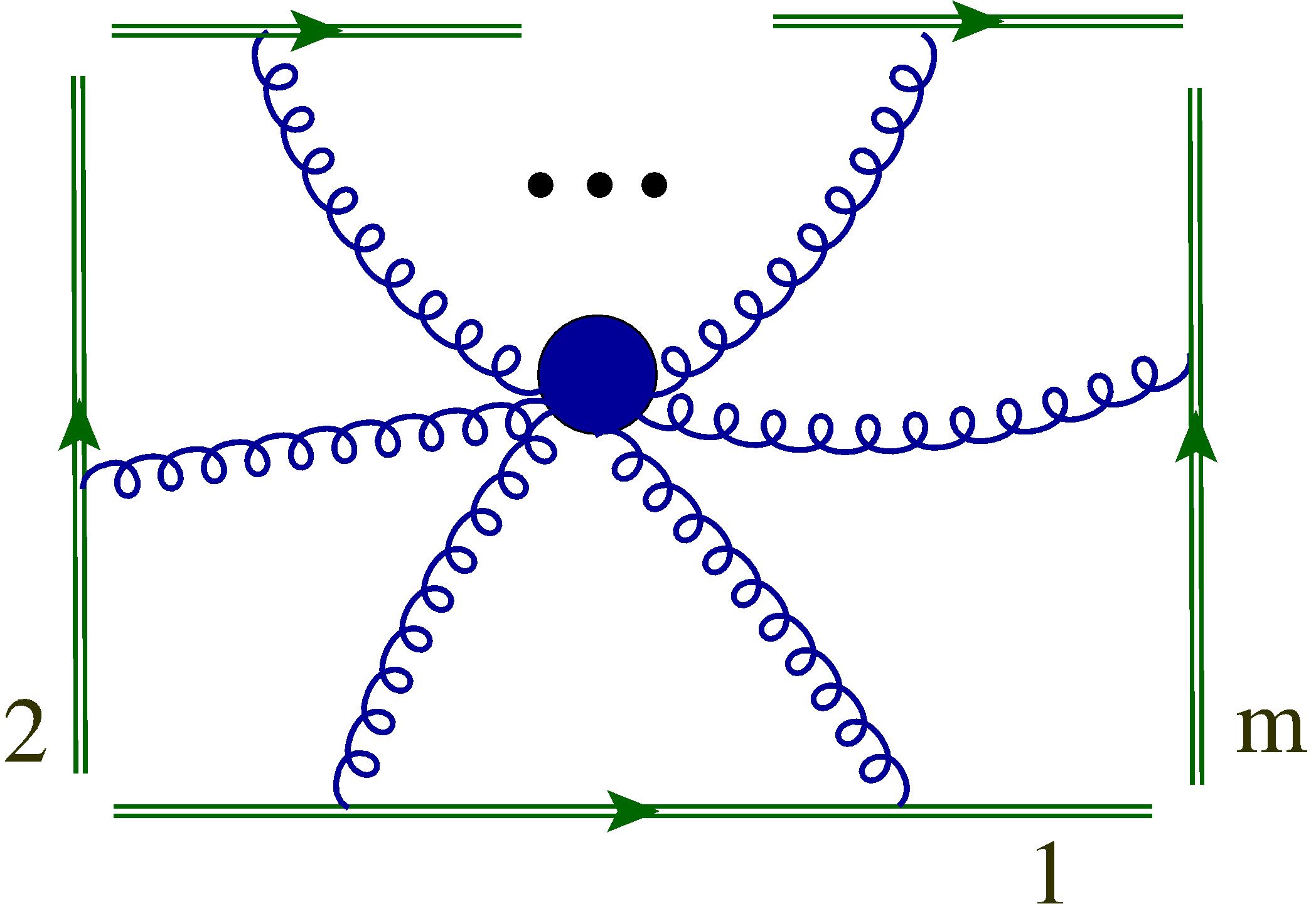} }
	\caption{ Equivalence of a general correlator containing a seagull vertex in diagram (a)  with an eikonal correlator in diagram (b)}
	\label{fig:SCorres1}
\end{figure}

Independence of the shuffle pattern from the type of vertices involved
implies the following. {For determining the shuffle of attachments in a Cweb, an \( n \)-point correlator connecting \( m \) lines in an NE Cweb with a seagull vertex behaves exactly the same as an \( n \)-point gluon correlator connecting the same lines in the eikonal case, as illustrated in Fig.~\eqref{fig:SCorres1}}. It is important to note that this equivalence holds up to the fact that the perturbative orders of the correlators may differ in certain cases.\footnote{In Fig.~\eqref{fig:SCorres1} both the diagrams have $ n $-point correlators, however, the the order of diagrams differ by $ g $ as seagull vertex in diagram (a) connecting to Wilson line 1 contributes $ \mathcal{O}(g^2)$ unlike any of the eikonal attachments in diagram (b) contributing a $ \mathcal{O}(g) $.} 

\begin{figure}[b]
	\centering
	\subfloat[][$\textbf{W}_{3}^{(1,1)}(2_s,2,1)$]{\includegraphics[scale=0.06]{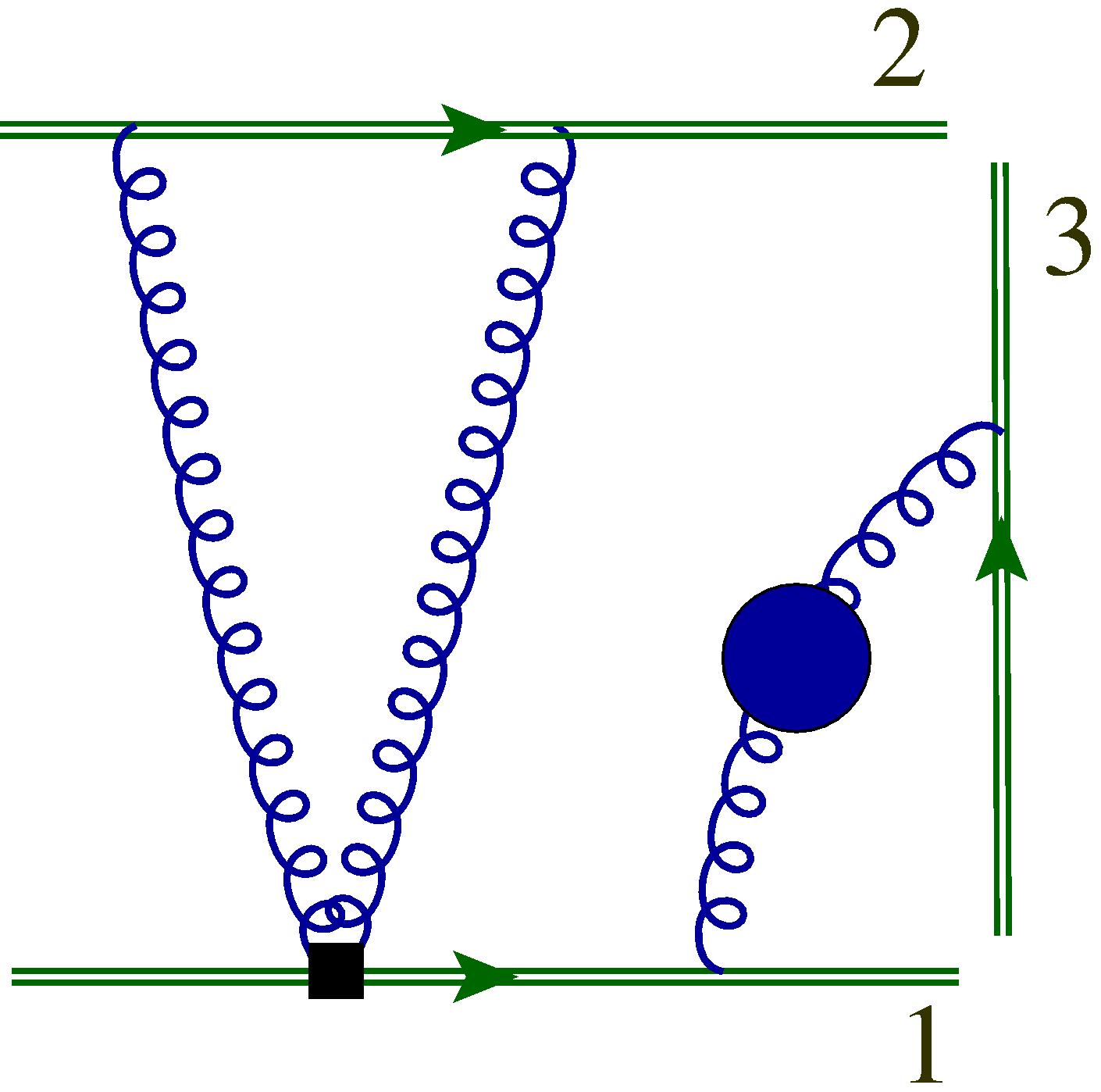} }
	\qquad 
	\subfloat[][ $\textbf{W}_{3}^{(1,1)}(2,2,1)$]{\includegraphics[scale=0.06]{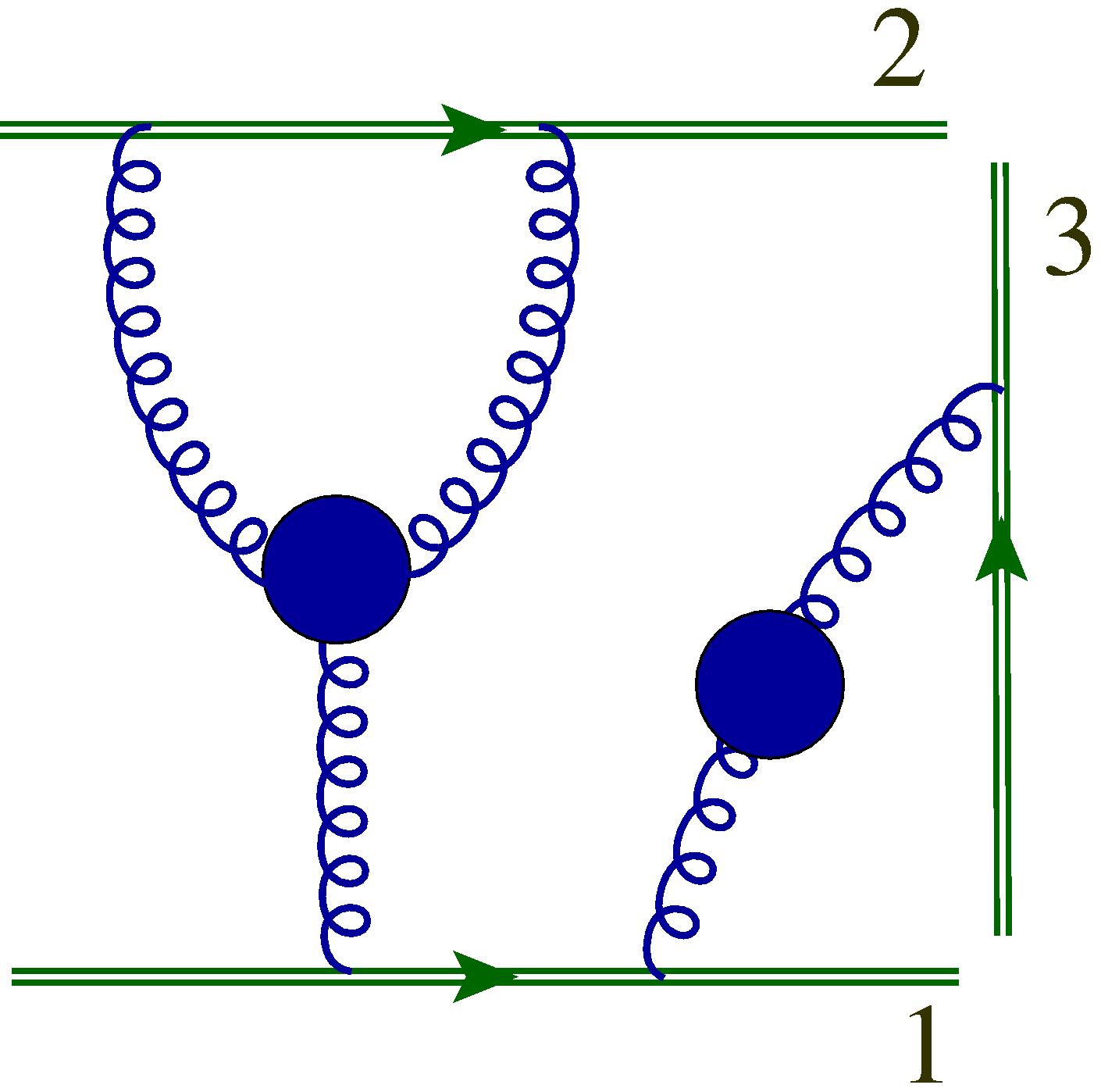} }
	\qquad
	\subfloat[][$\textbf{W}_{3}^{(1,1)}(2,\overline{1},1)$]{\includegraphics[scale=0.06]{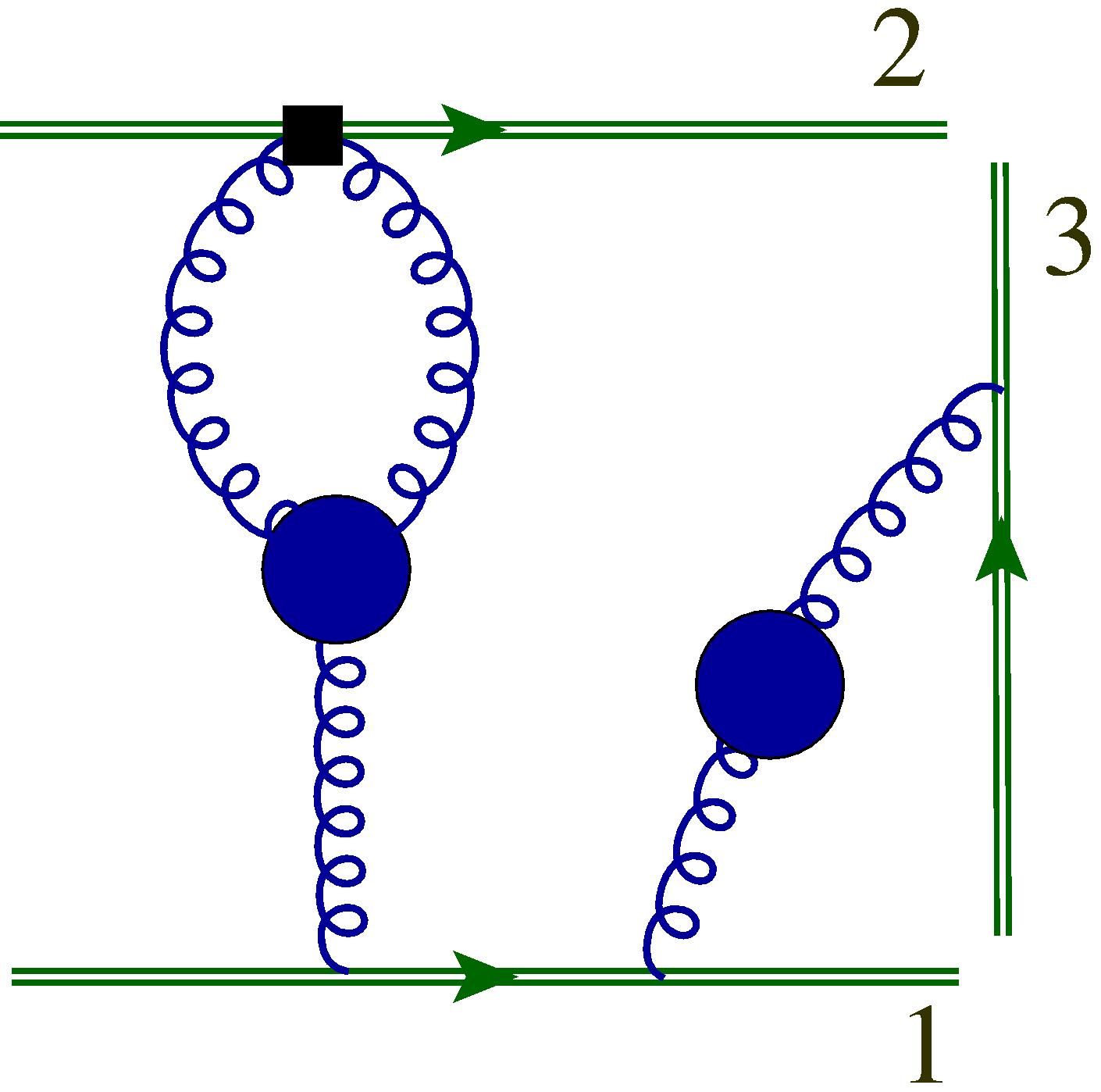} }
	\qquad
	\subfloat[][$\textbf{W}_{3}^{(2)}(2,1,1)$]{\includegraphics[scale=0.06]{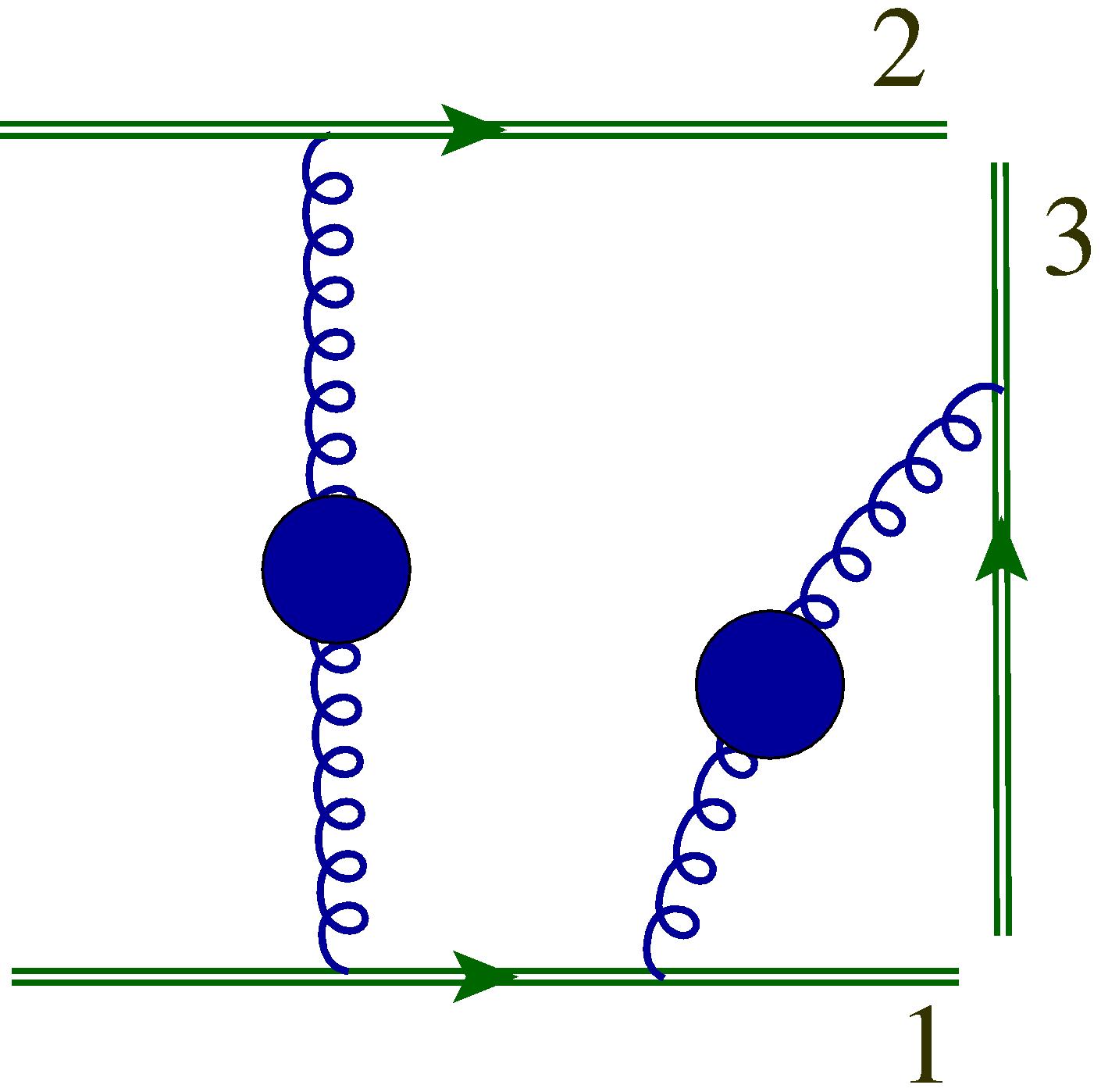} }
	\caption{NE Cwebs (a) and (c) containing a three-point and a two-point correlators with a seagull vertex. The corresponding eikonal correlators in eikonal Cwebs (b) and (d).}
	\label{fig:SeagullCorres1}
\end{figure}

Two NE Cwebs depicted in Fig.~\eqref{fig:SeagullCorres1} that appear at three-loop illustrate the above statements. The shuffling of correlator attachments in the NE Cwebs in Fig.~\eqref{fig:SeagullCorres1}a and Fig.~\eqref{fig:SeagullCorres1}c are same to those in Fig.~\eqref{fig:SeagullCorres1}b and Fig.~\eqref{fig:SeagullCorres1}d, respectively. All four of these Cwebs have two diagrams, as the shuffle is restricted to line 1. The difference in perturbative order is evident in the NE Cweb shown in Fig.~\eqref{fig:SeagullCorres1}c, where the associated eikonal Cweb in Fig.~\eqref{fig:SeagullCorres1}d contributes at two loops instead of three.

\subsection{Identicality of posets}\label{sec:HasseUniqueness}

 Here we revisit the concept of the symmetry factor for a diagram  from the perspective of order theory in mathematics~\cite{Dukes:2013gea} and generalise it for NE Cwebs.
%
% Symmetry factor \( s(D) \) for a diagram \( D \) is a key entity that will be used throughout this paper. \\
%	
%\noindent \textbf{Symmetry factor ($ s $-factor): }	 The symmetry factor $ s(d) $ for a diagram $ d $ is defined as the number of
%	ways in which the gluon correlators can be sequentially shrunk to their common origin. \\
%
In the case of eikonal webs, leading UV singularities of the multi-eikonal vertex belonging to the diagrams of a web are equal up to the factor $ s(D) $~\cite{Gardi:2011yz}. On the other hand, later in the literature, $ s $-factor has also been studied through the concept of posets and Hasse diagrams, which provides its another independent description. A connection between the \( s \)-factor and Hasse diagrams was established in~\cite{Dukes:2013gea} which we briefly review here.

A partially ordered set (poset) is a set 
$ P $ with a partial order 
$ \leq $ defined among the elements of the set. A relation is called a partial order if for all $ a,\,b,\,c \in P $, the following are satisfied:
\begin{enumerate}
\item  Reflexivity: $ a \leq a $,
\item  Antisymmetry: if $ a \leq b $ and $ b \leq a $ then $ a = b $,
\item  Transitivity: if $ a \leq b $ and $ b \leq c\,$ then $ a \leq c $.
\end{enumerate}
A graphical representation of a given poset is called a Hasse diagram. In these diagrams, each element of \( P \) is represented by a vertex, and a line is drawn between two distinct elements \( a \) and \( b \) with the element \( a \) placed below \( b \) if \( a \leq b \).

\begin{figure}[h!]
  	\centering
  	\subfloat[][]{\includegraphics[scale=0.06]{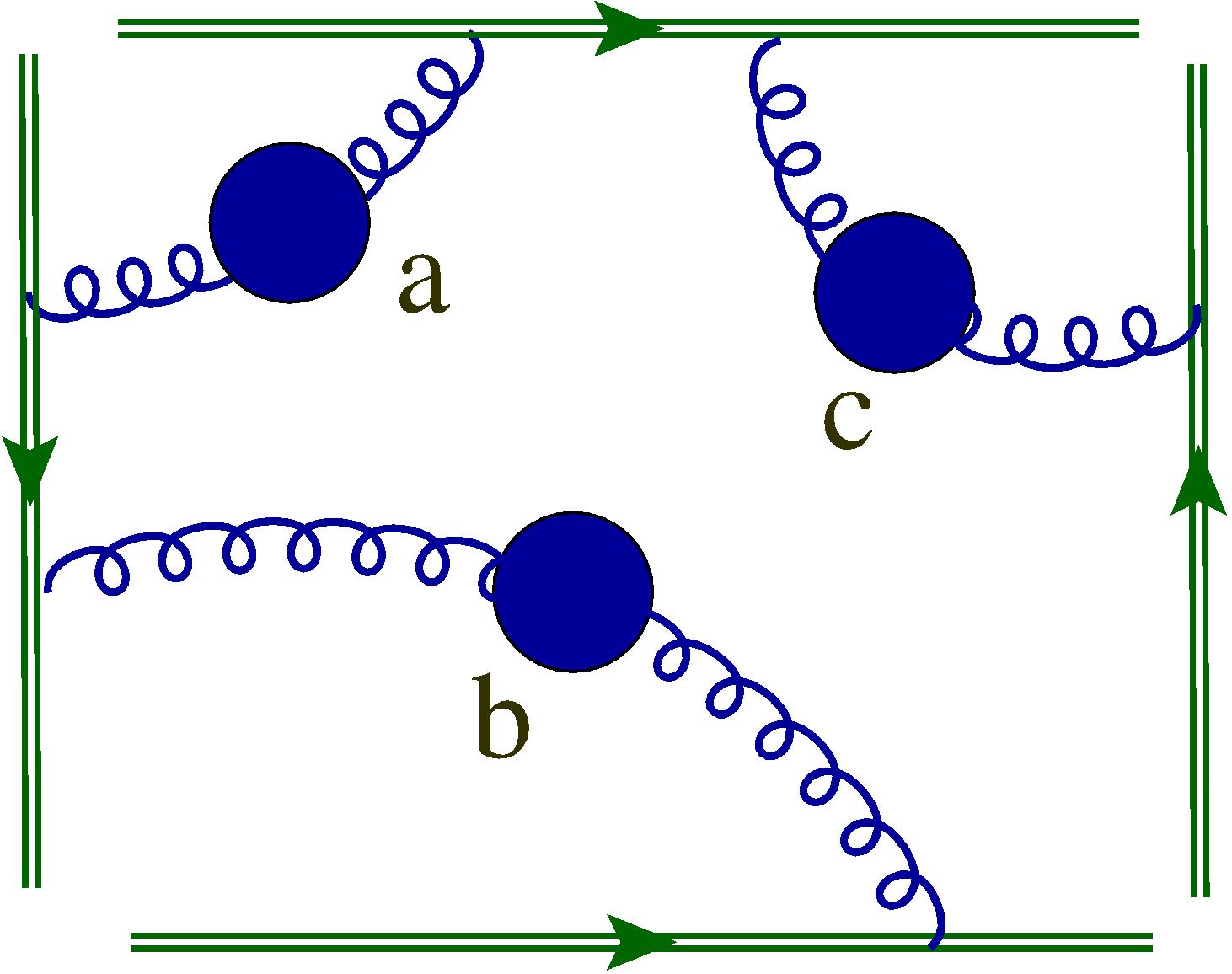} }
  	\hspace{1.5cm}
  	\subfloat[][ ]{\includegraphics[scale=0.09]{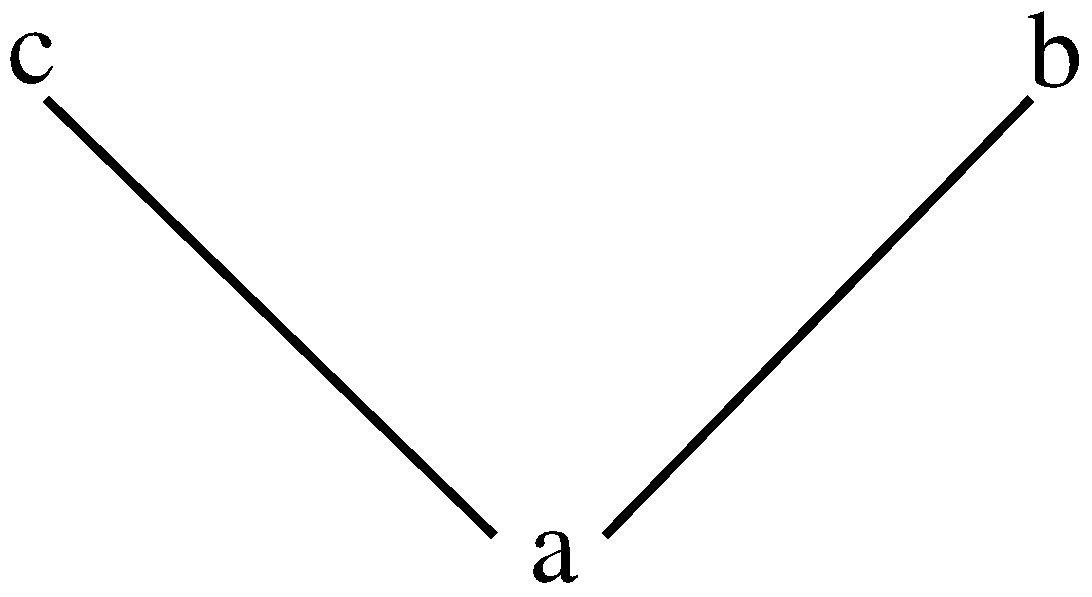} }
  	\hspace{1.5cm}
  	\subfloat[][]{\includegraphics[scale=0.09]{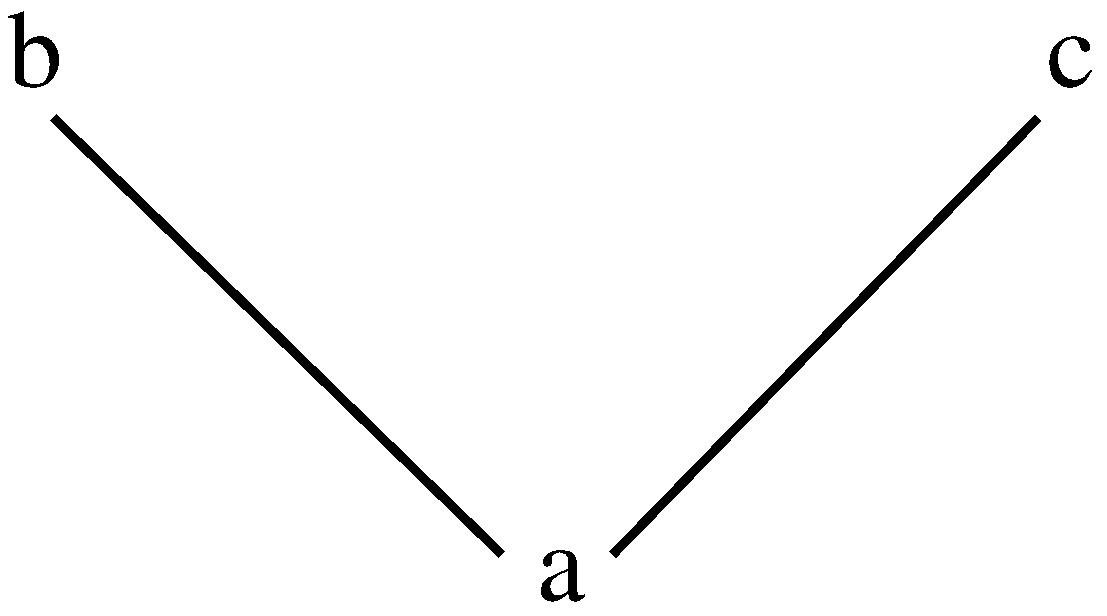} }
  	\caption{Examples of linear extension and Hasse diagrams}
  	\label{fig:hasseExamples}
  \end{figure}

We associate a poset with a 
 diagram $D$ of a Cweb such that its elements correspond to the correlators present in $ D $, where all the correlators that are entangled contribute as only one element to the poset.
 For any two such correlators $ a $ and $ b $, if $ a $ is closer  than $ b $ to the origin of Wilson line, then $a\leq b$.
  For the poset $ P =\{a,b,c\} $, with the relations \( a \leq b \), \( a \leq c \) that corresponds to the diagram shown in Figure~\eqref{fig:hasseExamples}a, we get two Hasse diagrams as shown in  Figures~\eqref{fig:hasseExamples}b and~\eqref{fig:hasseExamples}c.

A partial order 
$ \leq^* $ on a set 
$ P $ is called an extension of another partial order 
$ \leq $ on 
$ P $ provided that for all elements 
$ {a,b\in P} $ whenever 
$ {a\leq b,} $ it is also the case that 
$ a\leq^* b $. A linear extension is an extension that is also a linear (total) order. A total order is a partial order with the property that every
pair of elements are related. 
The symmetry factor $  s $ of a diagram \( D \) is defined to be the number of linear extensions of the poset \( P \) associated with \( D \), it also gives the number of distinct ways to shrink the correlators independently to the origin~\cite{Dukes:2013gea}.
  
  \vspace{0.5cm}

Identicality of shuffle implies the identicality of posets, i.e. the poset of an NE and an eikonal Cweb are the same if an NE correlator is replaced by an eikonal correlator as described in the section~\ref{sec:EandNEa}. 
Therefore the $ s$-factors of the diagrams in a NE Cweb can be determined through the number of linear extensions.

\vspace{0.5cm}

We conclude that the definition of the $ s $-factor and thus the column weight vector $ S $ remains unchanged as we go from eikonal to NE Cwebs. We next implement the formalism of~\cite{Agarwal:2022wyk} to determine the Cweb mixing matrices for NE Cwebs at the three-loop order.

\begin{figure}[t]
	\centering
	\subfloat[][$\textbf{W}_{3}^{(3)}(3,2,1)$]{\includegraphics[scale=0.06]{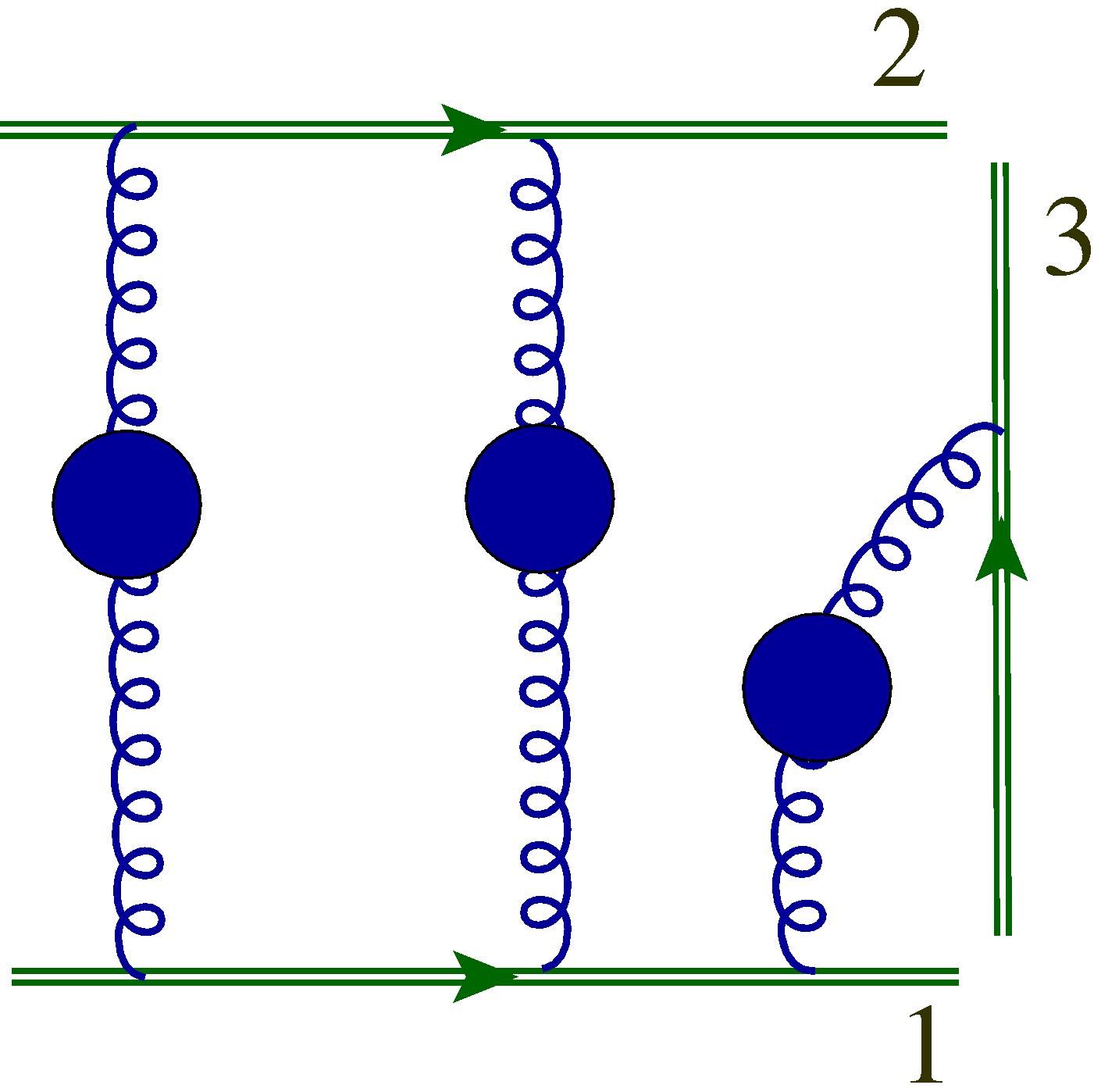} }
	\qquad \hspace{1cm}
	\subfloat[][ $\textbf{W}_{3, \text{I}}^{(1,1)}(2_s,2,1)$]{\includegraphics[scale=0.06]{SeagullAL2a} }
	\qquad \hspace{1cm}
	\subfloat[][$\textbf{W}_{3,\text{II}}^{(1,1)}(2_s,2,1)$]{\includegraphics[scale=0.06]{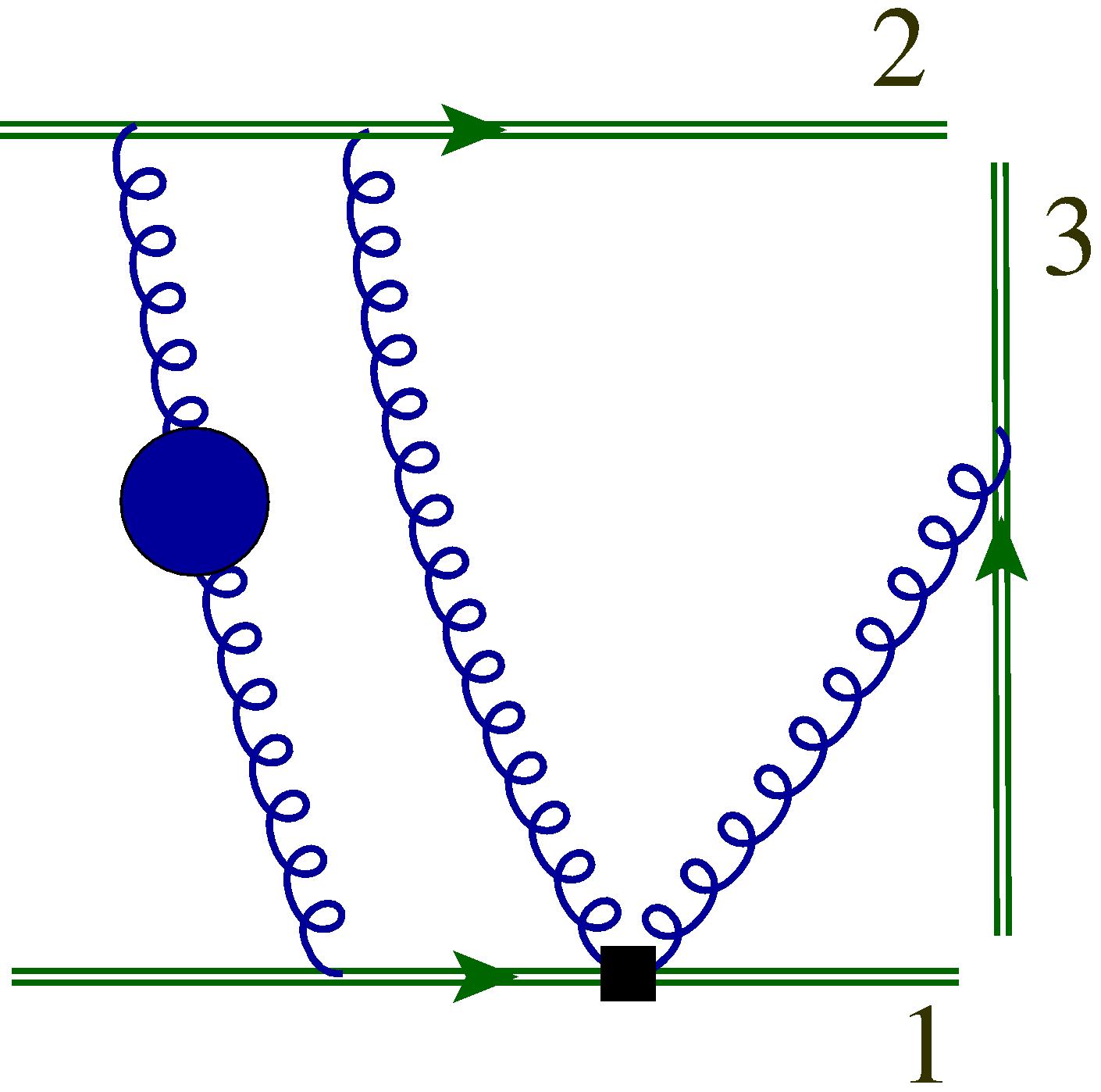} }
	%	\qquad
	%	\subfloat[][$\textbf{W}_{3}^{(1,1),\text{NE}}(1,2,2)$]{\includegraphics[scale=0.06]{SeagullALc} }
	\caption{ A three loop eikonal Cweb (a) that generates NE Cwebs (b) and (c) using step one and two to incorporate seagull vertex, at three loops.}
	\label{fig:SeagullAlgo2}
\end{figure}

\section{Direct Construction of NE Cwebs}\label{sec:DNE}
We begin this section by generating the NE Cwebs at three loops. The NE Cwebs at a given perturbative order can be obtained from eikonal Cwebs at the same order by shrinking the propagator between any two attachments on the same Wilson line into a seagull vertex (see appendix~\ref{sec:NEfeynmanrules}). There are 19 NE Cwebs present at three loops which have a seagull vertex; out of which nine are fully connected ($ S=\{1\} $), provided in appendix~\ref{sec:ConnectedWebs}, each having a trivial mixing matrix, \textit{i.e.}, 
$ R=1 $. The remaining 10 Cwebs belong to two groups: (a) one for which all the diagrams have $s \neq 0$, and
(b) those with at least one zero 
$ s $-factor. Mixing matrices for the first group can be directly determined using the Uniqueness theorem~\cite{Agarwal:2022wyk}. A careful observation reveals that the second group of NE Cwebs belongs to the Octopus Pair Cwebs, given in section (4.2) of~\cite{Agarwal:2022wyk}. 

 Before presenting the results, let us first explain the nomenclature of NE Cwebs; 
we follow the definition of Cweb given in \cite{Agarwal:2020nyc,Agarwal:2021him}. Cwebs are not fixed-order quantities but instead have their own expansion in powers of the gauge coupling \( g \). 
We will use the notation \( W_n^{(c_2, \ldots, c_p)} (k_1, \ldots, \overline{k}_{j}, \ldots, k_n) \) to describe a NE Cweb. This is constructed from \( c_m \) \( m \)-point connected gluon correlators (\( m = 2, \ldots, p \)), where one of the correlators includes a seagull vertex. 
 Here, \( k_i \) refers to the number of attachments on the \( i^{\text{th}} \) Wilson line, where bar on the \( \overline{k}_{j} \)
 explicitly tells that out of $ k_j $ attachments one is a seagull vertex on the Wilson line.
Since an \( m \)-point connected gluon correlator starts at \( {\cal O}(g^{m - 2}) \), and each attachment to a Wilson line adds one more power of \( g \), the expansion for a Cweb can be written as

\begin{align}
W_n^{(c_2, \ldots , c_p)} (k_1,\ldots, \overline{k}_{j}, \ldots  , k_n)  \, &= \, 
g^{\, (\overline{k}_{j}+1)\,+\overline{\sum}_{i = 1}^n k_i\, \, + \, \sum_{r = 2}^p c_r (r - 2)} \nonumber \\ 
  & \qquad	 \, \times \sum_{j = 0}^\infty \,W_{n, \, j}^{(c_2, \ldots , c_p)} (k_1,\ldots, \overline{k}_{j}, \ldots  , k_n) \, g^{2 j} \, ,
\label{pertCweb}
\end{align}
which defines the perturbative coefficients $W_{n, \, j}^{(c_2, \ldots , c_p)} 
(k_1, \ldots  , k_n)$. The term \( \overline{k}_{j} + 1  \) in the exponent of \( g \) in the prefactor equals the total number of eikonal attachments and the contribution from a seagull attachment to the $ j^{th} $ Wilson line. Also the $ \overline{\sum} $ represents a summation over $ 1\leq i\leq n $ excluding $ i=j $.
Now we present the details of calculations in the following sections.

\subsection{Cwebs with $ S=\{s\neq 0 \} $}

\begin{figure}[h]
	\centering
				\includegraphics[scale=0.06]{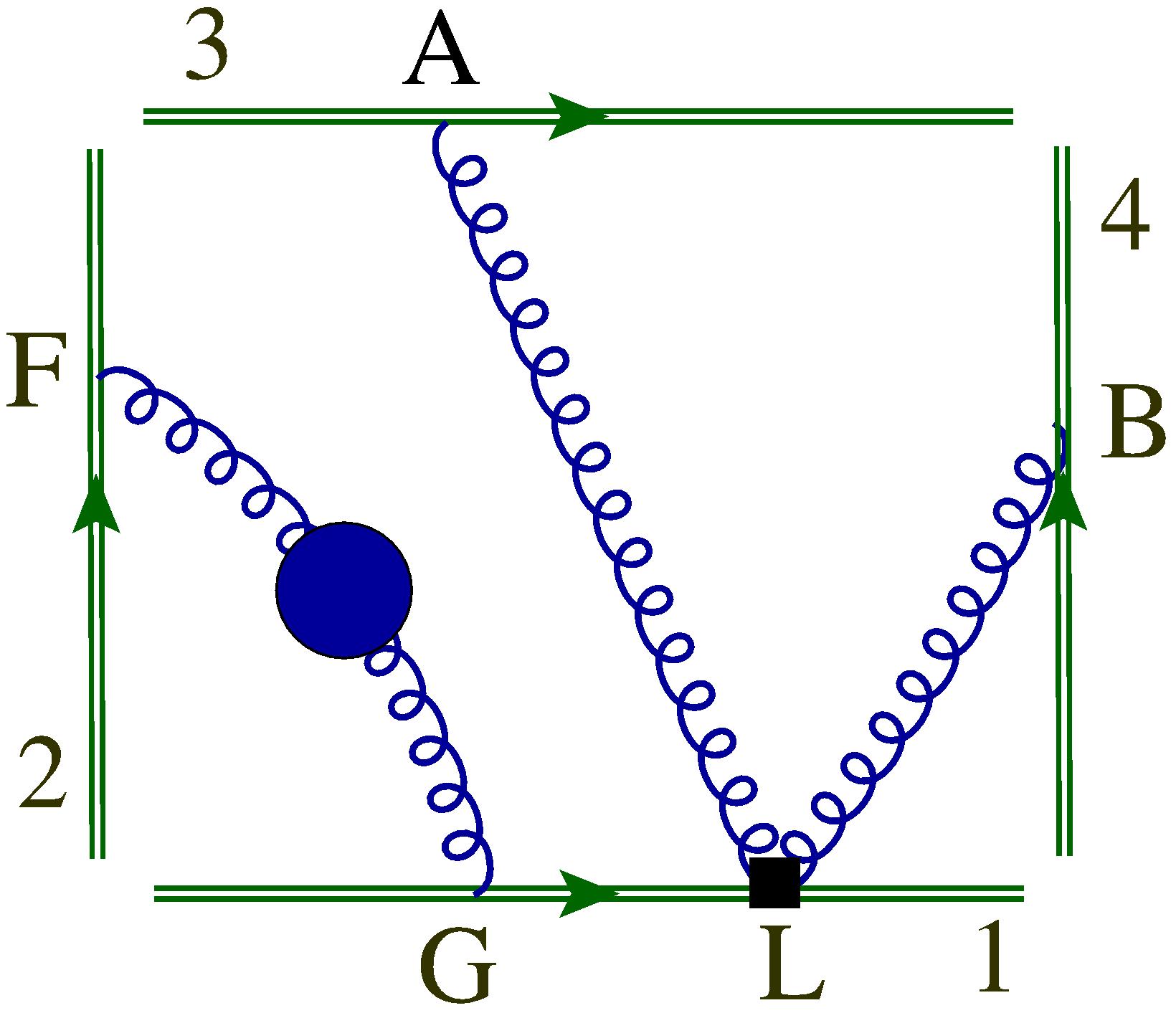}
				\caption{Diagram $ C_1 $ of NE Cweb belongs to the first class}
				\label{fig:NEcwebC1}
\end{figure}

At three loops there are 5 such NE Cwebs whose column weight vectors do not contain any zero entry. The column weight vector for each of these five Cwebs is same and is,
\begin{align}
	S=\{1,1\}\equiv\{1_2\}\,,
\end{align}
where $ \{\ldots, 1_2\}=\{\ldots, 1,1\} $.
The mixing matrix associated with this $ S $ has been directly constructed in~\cite{Agarwal:2021him,Agarwal:2022wyk} and is,
 \begin{align}\label{eq:R(1_2)}
 	R = R(1_2) \equiv \left(\begin{array}{cc}
 		\frac{1}{2}  & -\frac{1}{2}\\
 		-\frac{1}{2} & \frac{1}{2}
 	\end{array}\right)\,,\quad \textrm{rank}(R(1_2))=1\,.
 \end{align}
It follows from the Uniqueness theorem that all of these five Cwebs have the mixing matrix given above.
Since the rank of the mixing matrix is one,
each of the Cwebs in this class has one independent exponentiated colour factor. One of the NE Cwebs $\textbf{W}_{4, \, \text{I}}^{(1,1)}(\overline{2},1,1,1)$ that belongs to this class is shown in Fig.~\eqref{fig:NEcwebC1}. The shuffles associated with the two diagrams can
be labelled by giving the sequences of the attachments, in the orderings defined by the orientation
of the Wilson lines. 
	This Cweb has two diagrams $ C_1 $ and $ C_2 $ with the order of attachments are $ \lbrace GL\rbrace$  and $ \lbrace LG\rbrace$ respectively. 
The exponentiated colour factor is
\begin{eqnarray}
	(YC)_1 = - if^{ack}  \sbk 1 \tb 2 \ta 3 \tc 4 
	- if^{bck} \ska 1 \tb 2 \ta 3 \tc 4\,. 
\end{eqnarray} 
The ECFs for each NE Cweb of this class is provided in appendix~\ref{sec:Direct-C-NEappendix}.  
\begin{figure}[b]
	\centering
	\includegraphics[scale=0.1]{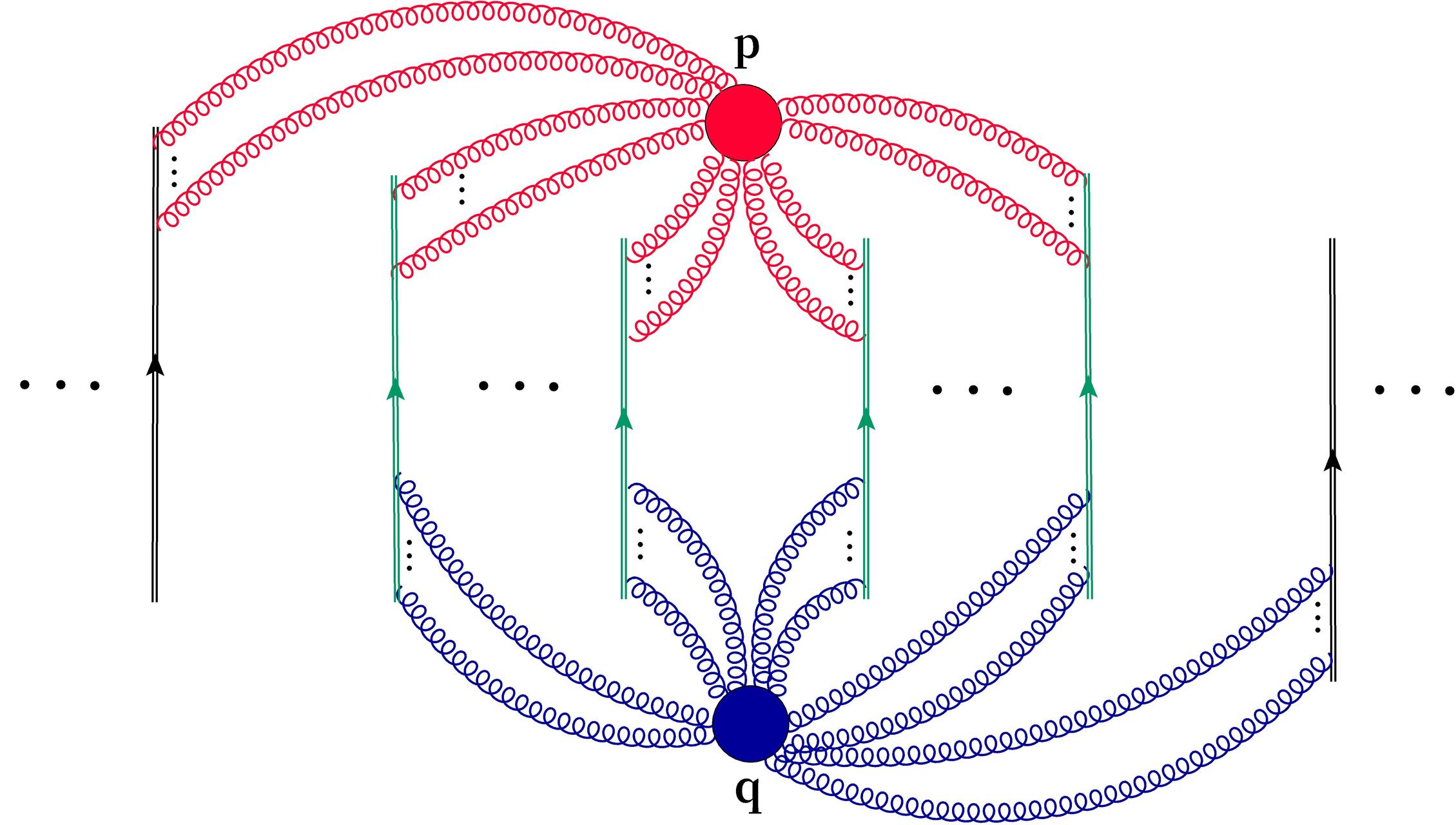}
	\caption{Octopus pair with two distinct $ p $ and $ q $ point correlators.}
	\label{fig:Octopus}
\end{figure}

\subsection{Cwebs with $ S= \{0,0,\cdots, 1_2 \} $ }

\begin{figure}
	\begin{minipage}{0.4\textwidth}
	\centering		\includegraphics[scale=0.06]{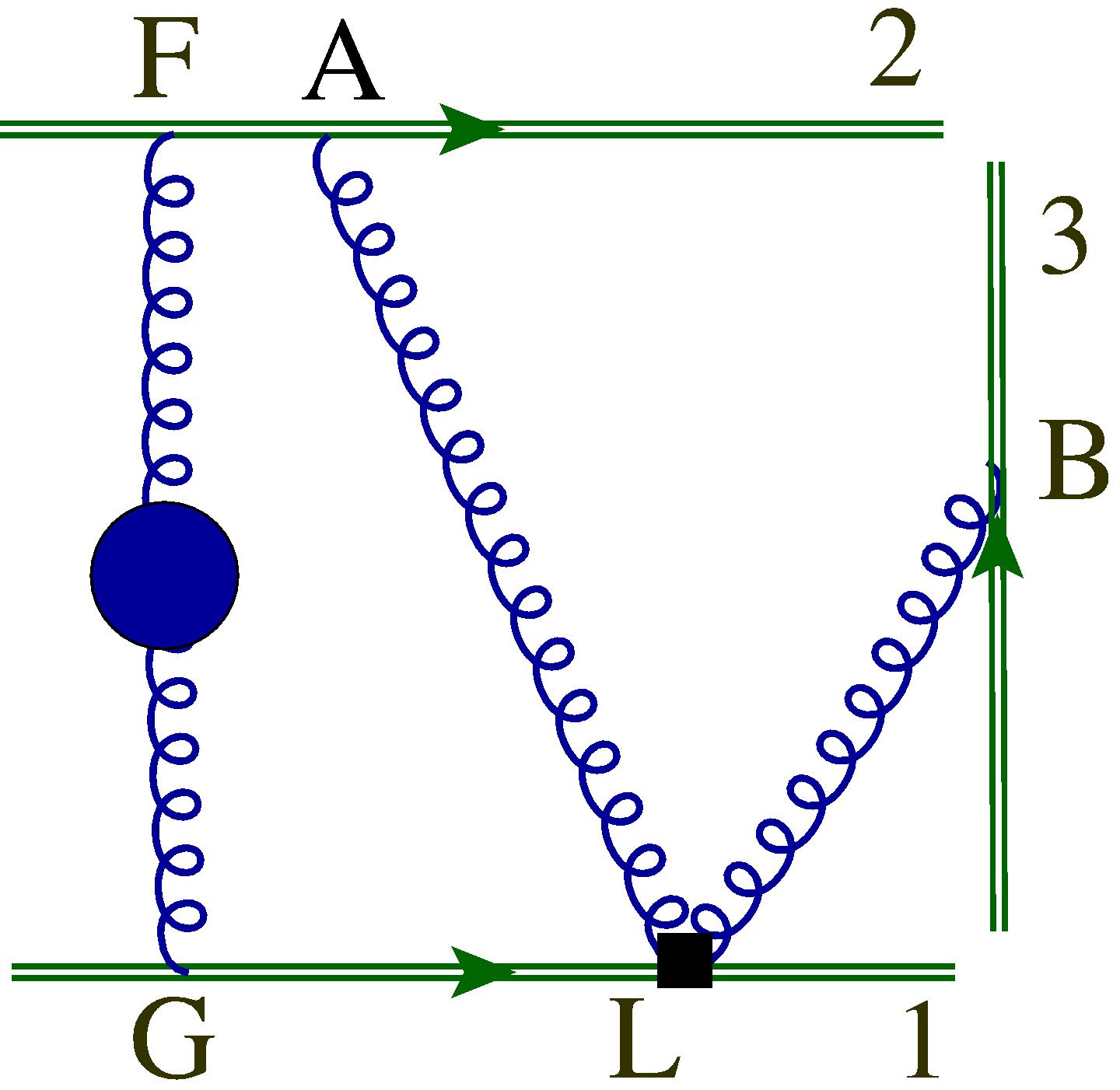}
	\end{minipage} \hspace{1cm}
	\begin{minipage}{0.4\textwidth}
		\begin{tabular}{ | c | c | c |}
			\hline
			\textbf{Diagrams} & \textbf{Sequences} & \textbf{S-factors} \\ \hline
			$C_1$ & $\{\lbrace GL \rbrace, \lbrace AF\rbrace\}$ & 0\\ \hline
			$C_2$ & $\{\lbrace LG \rbrace, \lbrace FA\rbrace\}$ & 0\\ \hline
			$C_3$ & $\{\lbrace GL\rbrace, \lbrace FA\rbrace\}$ & 1 \\ \hline
			$C_4$ & $\{\lbrace LG \rbrace, \lbrace AF\rbrace\}$ & 1\\ \hline
		\end{tabular}	
	\end{minipage}
\caption{The diagram $ C_3 $ of NE Cweb and the sequence of attachments for all four diagrams}
\label{fig:Octopus-Example}
\end{figure}

The last remaining 5 Cwebs belong to this category. The column weight vector for this class has following form 
\begin{align}
	S=\{ 0,0,\cdots, 1_2 \}\,.
\end{align}
The Cwebs of this category are referred to as \textit{Octopus Pair} Cwebs in~\cite{Agarwal:2022wyk}. These Cwebs are composed of two distinct gluon correlators, as illustrated in Fig.~\ref{fig:Octopus}. For a general Cweb in this category containing \( n \) diagrams, the mixing matrix takes the following form~\cite{Agarwal:2022wyk}:
\begin{align}
	R_{n \times n} = \left( \begin{array}{cc}
		I_{(n-2) \times (n-2)} & B_{(n-2) \times 2}\\
		O_{2\times (n-2)} & R(1_2)
	\end{array}\right)\equiv R(0_{n-2},1_2)\,\,.
\end{align}
Here, \( I_{(n-2) \times (n-2)} \) and \( O_{2 \times (n-2)} \) denote the identity and null matrices, respectively and  $ R(1_2) $ is a already known basis matrix~\cite{Agarwal:2022wyk}, with the form given in eq.~\eqref{eq:R(1_2)}. Each element of the matrix \( B \) is \( -1/2 \). Additionally, the rank of the mixing matrix is \( n-1 \), indicating that these Cwebs have \( n-1 \) independent exponentiated color factors.
Next, we present the mixing matrix and its rank for  NE Cweb  $\textbf{W}_{3, \, \text{II}}^{(1,1)}(\overline{2},2,1)$ of this class.
	This Cweb has four diagrams, one of which is displayed in Fig.~\eqref{fig:Octopus-Example}. The table in Fig.~\ref{fig:Octopus-Example} gives
	the chosen order of the four shuffles of the gluon attachments, and the corresponding
	$s$-factors.\\
	
	\noindent The mixing matrix and its rank is given by
	\begin{align}
		R=R(0_2,1_2)    \,\qquad;	\qquad 
		r(R(0_2,1_2)) = 4-1 = 3  \, .
	\end{align}
The exponentiated colour factors are
\begin{align*}
	(YC)_1 &=
	- i f^{acn}\sba 1 \tc 1 \tb 2 \tn 3  
	- i f^{ack}\sbk 1 \tb 2 \tc 3 \ta 3 
	- i f^{bck} \ska 1 \tb 2 \tc 3 \ta 3 
	\\
	(YC)_2 &=
	-i f^{ack} \sbk 1 \tb 2 \ta 3 \tc 3 
	-i f^{bck} \ska 1 \tb 2 \ta 3 \tc 3 \\
	(YC)_3 &=
	-i f^{acn} \sba 1 \tc 1 \tb 2 \tn 3 
\end{align*} 
The results for remaining NE Cwebs for this class is provided in the appendix~\ref{sec:Direct-C-NEappendix}.  All three-loop NE Cwebs, their mixing matrices, and rank are summarised in table~\eqref{tab:non-prime-singlet2}.

\begin{table}[h]
	\begin{center}
		\begin{tabular}{|c|l|c|c|}
			\hline 
			& &  &\\
			Column Weight Vector  &$ \quad \quad$ Cweb & Mixing matrix $ R $ & rank r(R) \\
			\hline  
			& &	&	\\
			&  $\textbf{W}_{4}^{(1,1)}(\overline{2},1,1,1)$ &  &\\
			\cline{2-2}
			&&&\\
			&  $\textbf{W}_{4}^{(1,1)}(\overline{1},1,2,1)$ & &\\
			\cline{2-2}
			&&&\\
			&$\textbf{W}_{3,\text{I}}^{(1,1)}(\overline{2},2,1)$ & &\\
			\cline{2-2}
			\{1, 1\} & 						& $ R(1_2)$ & $1$\\
			&  $\textbf{W}_{3}^{(1,1)}(\overline{2},1,1)$ &  &\\
			\cline{2-2}
			& 							&	&\\
			& $\textbf{W}_{3}^{(1,1)}(\overline{1},2,1)$ & &\\
			\cline{2-2}
			\hline
			&     &        &       \\
			&  $\textbf{W}_{3, \, \text{II}}^{(1,1)}(\overline{2},2,1)$ & &\\
			\cline{2-2}
			\{0,0,1,1\}& 									& $ R(0_2,1_2)$ & $3$\\
			& $\textbf{W}_{3}^{(1,1)}(\overline{1},2,2)$& &\\
			\cline{2-2}
			& 									&	&\\
			& $\textbf{W}_{2}^{(1,1)}(\overline{2},2)$ & &\\
			\hline
			& 							&  &\\
			\{0,1,1\}	& $\textbf{W}_{3}^{(1,1)}(\overline{1},3,1)$& $ R(0_1,1_2)$ & 2\\
			\hline 
			& & &\\
			\{0,0,0,0,1,1\}	&$\textbf{W}_{2}^{(1,1)}(\overline{2},3)$&  $ R(0_4,1_2) $& 5 \\
			\hline 	
		\end{tabular}
	\end{center}
	\caption{NE Cwebs at three loops and their mixing matrices.}
	\label{tab:non-prime-singlet2}
\end{table}

\section{Summary}\label{sec:results}

In the last decade, there has been a growing interest in studying the scattering amplitudes and cross-sections beyond the eikonal (soft) approximation. The subleading corrections are known as next-to-eikonal or next-to-soft corrections. In this paper we consider the diagrams that exponentiate in the next-to-eikonal limit, which are formally known as next-to-eikonal Cwebs. We establish two correspondences between eikonal and next-to-eikonal Cwebs via posets to show that the mixing matrices of NE Cwebs can be obtained using the concepts developed for eikonal Cwebs in~\cite{Agarwal:2022wyk}.
 The study of the kinematic of NE Cwebs is a quite interesting domain that can further shed light on the fact if there exist kinematic relations between  eikonal Cwebs and NE Cwebs. 

\section*{Acknowledgements}

SP, AT would like to thank Neelima Agarwal and Lorenzo Magnea for collaborating on the earlier projects. AD would like to thank CSIR, Govt. of India, for an SRF fellowship (09/1001(00 49)/2019- EMR-I). The research of SP is 
supported by the SERB Grant SRG/2023/000591 and NISER Plan Project No.: \lq \lq RIN-4001\rq \rq. AS would like to thank CSIR, Govt. of India, for an SRF fellowship (09/1001(0075)/2020-EMR-I).

\appendix

\section{Effective Feynman rules}\label{sec:NEfeynmanrules}
The NE effective Feynman rules have been derived from the path integral approach~\cite{Laenen:2008gt} for complex scaler field coupled with abelian gauge field, and for the fermions interacting with non-abelian gauge field with the diagrammatic approach~\cite{Laenen:2010uz}. The eikonal as well as next-to-eikonal effective vertices that contribute to the soft emission of non-abelian gauge bosons from fermion lines are following:
\begin{figure}[h]
	\centering
	\subfloat[][]{\includegraphics[scale=0.5]{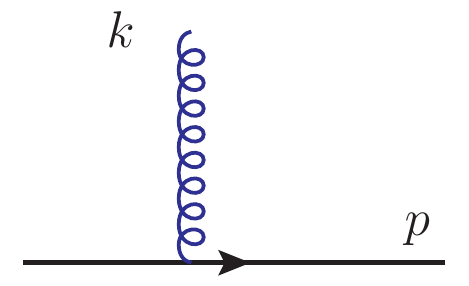} }
	\qquad \hspace{1cm}
	\subfloat[][]{\includegraphics[scale=0.5]{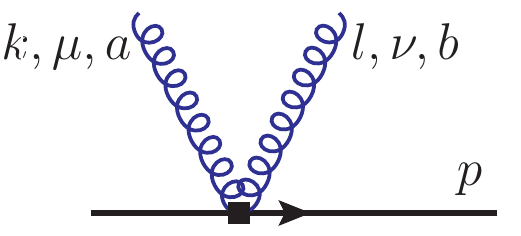} }
	\caption{(a) Single gluon next to eikonal emission vertex, (b) Two gluon next to eikonal emission vertex also called seagull vertex}\label{fig:Veikonal}
\end{figure}
\noindent In the eikonal approximation, only one type of vertex appears, corresponding to the emission of a single gluon, as shown in Fig.~(\ref{fig:Veikonal}\textcolor{blue}{a}). The Dirac structure of this vertex is:

\begin{align}
	{\bf T}^a_{i,j} \left(\frac{p^{\mu}}{p \cdot k}\right)
\end{align}

\noindent In the next-to-eikonal approximation, two types of vertices appear. The first is the same as in the eikonal approximation: the single gluon emission depicted in Fig.~(\ref{fig:Veikonal}\textcolor{blue}{a}). This vertex contributes at both the eikonal and next-to-eikonal levels, with the following NE contribution:

\begin{align}
	{\bf T}^a_{i,j} \left\{ - \dfrac{k^\mu}{2 p\cdot k} \; +\;  k^2\dfrac{p^\mu}{2(p\cdot k)^2} \;+\; \dfrac{ik_\nu \Sigma^{\mu\nu}}{p\cdot k} \right\}\,, \qquad  \Sigma^{\mu\nu} \equiv \dfrac{i}{4}[\gamma^\mu, \gamma^\nu]  
\end{align}
The contribution of two gluon emission vertex $ \mathbf{S}_{\mu \nu}^{ab} $, shown in Fig.~(\ref{fig:Veikonal}\textcolor{blue}{b}) known as seagull vertex is, 
\begin{align}\label{eq:SeagullFormula}
	{\bf S}_{\mu \nu}^{ab} = \{{\bf T}^a ,{\bf T}^b\} \dfrac{ g_{\mu \nu} }{p\cdot (k+l)} \;+ \frac{1}{2} \; \left[ {\bf T}^a ,{\bf T}^b \right] \dfrac{  [\gamma^\mu, \gamma^\nu]}{p\cdot (k+l)} \equiv A_{\mu\nu} { \{{\bf T}^a, {\bf T}^b\} }\,+\,B_{\mu\nu} \left[ {\bf T}^a, {\bf T}^b \right] 
\end{align}

\noindent The generators of gauge group follow $ SU(3) $ algebra, however, any encounter of commutation of seagull vertex with any single emission vertex is non trivial. The calculation for this --- suppressing the Lorentz indices ---  is as follows
\begin{align}\label{eq:SeagullCommutator}
	\begin{split}
		\left[ {\bf T}^a, {\bf S}^{bc} \right] &= \left[ {\bf T}^a, A { \{{\bf T}^b, {\bf T}^c\} }\,+\,B \left[ {\bf T}^b, {\bf T}^c \right] \right]\\
		&= if^{abk}{\bf S}^{ck}\;+\;if^{ack}{\bf S}^{kb}
	\end{split}
\end{align}
where the structure constants are arise from the $ SU(3) $ algebra of generators of gauge fields as
\begin{align}
	\left[ {\bf T}^a,  {\bf T}^b       \right] = if^{abc}{\bf T}^c\,.
\end{align}

\section{ECFs for NE Cwebs}\label{sec:Direct-C-NEappendix}
In this section we provide the mixing matrices, Diagonalizing matrix and ECFs of all the NE Cwebs present at three loop having a seagull vertex.

\subsection{Cwebs with $ S=\{s\neq 0 \} $}
 In the following we provide the NE Cwebs contributing at three loops order, belonging to this class.

\begin{itemize}
	\item[{\bf 1}.]  $\textbf{W}_{4, \, \text{I}}^{(1,1)}(\overline{2},1,1,1)$
	
	This Cweb has two diagrams, one of which is displayed below. The table gives
	the chosen order of the two shuffles of the gluon attachments, and the corresponding
	$S$ factors.
	\begin{minipage}{0.4\textwidth}
			\includegraphics[scale=0.06]{4Web1}
	\end{minipage} 
	\begin{minipage}{0.46\textwidth}
		\begin{tabular}{ | c | c | c |}
			\hline
			\textbf{Diagrams} & \textbf{Sequences} & \textbf{S-factors} \\ \hline
			$C_1$ & $\lbrace GL\rbrace$ & 1 \\ \hline
			$C_2$ & $\lbrace LG \rbrace$ & 1\\ \hline
		\end{tabular}	
	\end{minipage}

	\noindent The $R$, $Y$ and $D$ matrices are given by
	\begin{align}
		R=\left(
		\begin{array}{cc}
			\frac{1}{2} & -\frac{1}{2} \\
			-\frac{1}{2} & \frac{1}{2} \\
		\end{array}\right)\,, \qquad Y=\left(
		\begin{array}{cc}
			-1 & 1 \\
			1 & 1 \\
		\end{array}\right)\,,\qquad	\qquad 
		D = \D{1} \, .
		\label{eq:4Web1}
	\end{align}
	Finally, the exponentiated colour factors are
	\begin{eqnarray}
		(YC)_1 = - if^{ack}  \sbk 1 \tb 2 \ta 3 \tc 4 
		- if^{bck} \ska 1 \tb 2 \ta 3 \tc 4 
	\end{eqnarray} 
	\vspace{2mm}
	
	%%%%%%%%%%%%%%%%%%
	\item[{\bf 2}.]  $\textbf{W}_{4}^{(1,1)}(\overline{1},1,2,1)$
	
		This Cweb has two diagrams, one of which is displayed below. The table gives
	the chosen order of the two shuffles of the gluon attachments, and the corresponding
	$S$ factors.
	\begin{minipage}{0.4\textwidth}
		\includegraphics[scale=0.06]{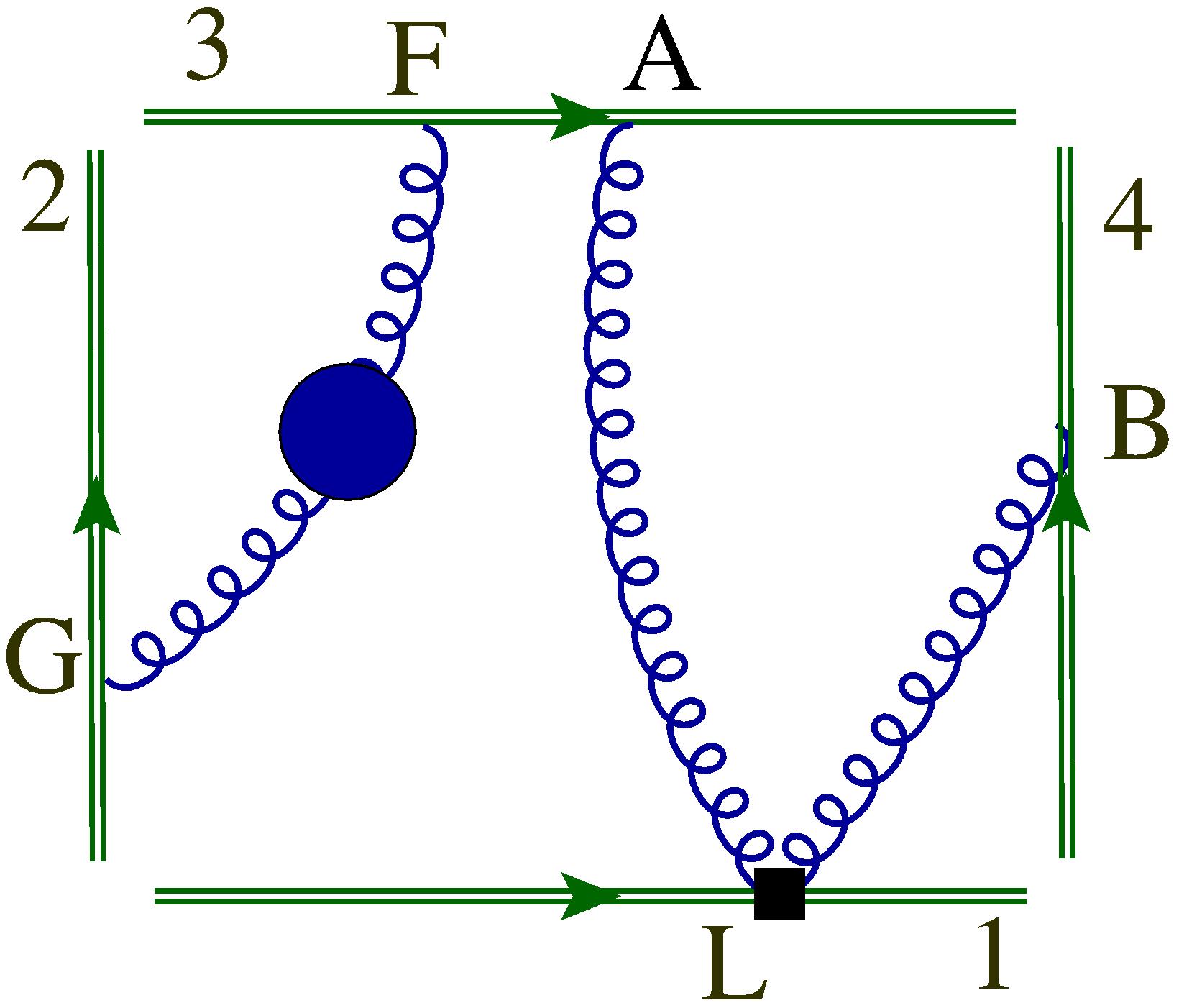}
	\end{minipage} 
	\begin{minipage}{0.46\textwidth}
		\begin{tabular}{ | c | c | c |}
			\hline
			\textbf{Diagrams} & \textbf{Sequences} & \textbf{S-factors} \\ \hline
			$C_1$ & $\{\lbrace FA\rbrace\}$ & 1 \\ \hline
			$C_2$ & $\{\lbrace AF \rbrace\}$ & 1\\ \hline
		\end{tabular}	
	\end{minipage}

	\noindent The $R$, $Y$ and $D$ matrices are given by
	\begin{align}
		R=\left(
		\begin{array}{cc}
			\frac{1}{2} & -\frac{1}{2} \\
			-\frac{1}{2} & \frac{1}{2} \\
		\end{array}\right)\,, \qquad Y=\left(
		\begin{array}{cc}
			-1 & 1 \\
			1 & 1 \\
		\end{array}\right)\,,\qquad	\qquad 
		D = \D{1} \, .
		\label{eq:4Web2}
	\end{align}
	Finally, the exponentiated colour factors are
	\begin{eqnarray}
		(YC)_1 =
		i\,f^{ack} \tk 1 \tc 2 \tb 4 \sba 3 
	\end{eqnarray}

	%%%%%%%%%%%%%%%%%%%%%%%%%%%%%%%%%%%%

	\item[{\bf 3}.]  $\textbf{W}_{3}^{(1,1)}(\overline{2},2,1)$
	
	This Cweb has two diagrams, one of which is displayed below. The table gives
the chosen order of the two shuffles of the gluon attachments, and the corresponding
$S$ factors.
\begin{minipage}{0.4\textwidth}
		\includegraphics[scale=0.06]{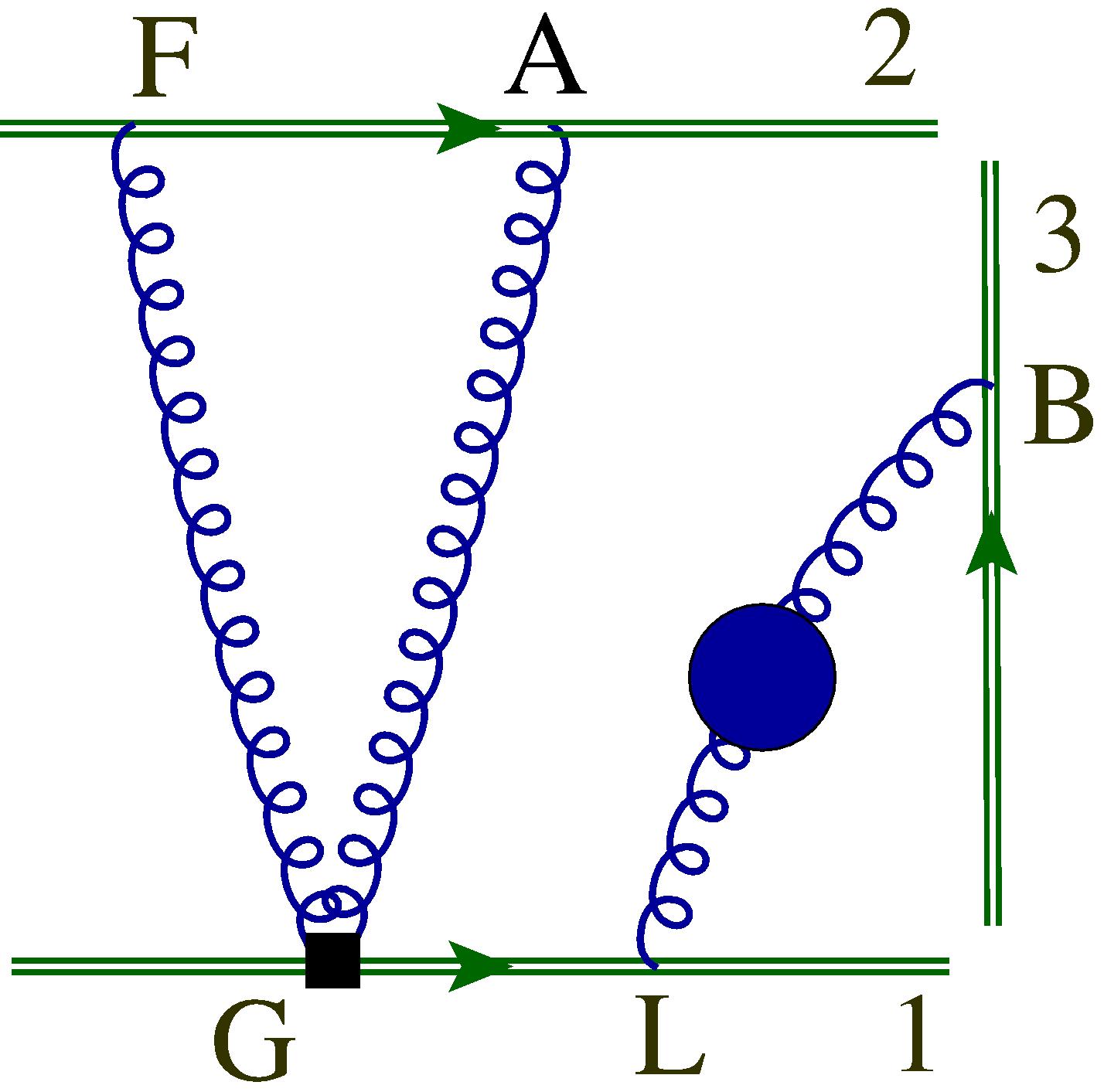}
\end{minipage} 
\begin{minipage}{0.46\textwidth}
	\begin{tabular}{ | c | c | c |}
		\hline
		\textbf{Diagrams} & \textbf{Sequences} & \textbf{S-factors} \\ \hline
		$C_1$ & $\{\lbrace LG\rbrace\}$ & 1 \\ \hline
		$C_2$ & $\{\lbrace GL \rbrace\}$ & 1\\ \hline
	\end{tabular}	
\end{minipage}

\vspace{0.5cm}
\noindent The $R$, $Y$ and $D$ matrices are given by
\begin{align}
	R=\left(
	\begin{array}{cc}
		\frac{1}{2} & -\frac{1}{2} \\
		-\frac{1}{2} & \frac{1}{2} \\
	\end{array}\right)\,, \qquad Y=\left(
	\begin{array}{cc}
		-1 & 1 \\
		1 & 1 \\
	\end{array}\right)\,,\qquad	\qquad 
	D = \D{1} \, .
	\label{eq:3Web3}
\end{align}
Finally, the exponentiated colour factors are
\begin{align*}
	(YC)_1 =
	- i f^{ack} \sbk 1 \tc 2 \tb 3 \ta 3  
	- i f^{bck} \ska 1 \tc 2 \tb 3 \ta 3  
\end{align*} 
	
	\item[{\bf 4}.]  $\textbf{W}_{3}^{(1,1)}(\overline{2},1,1)$
	
This Cweb has two diagrams, one of which is displayed below. The table gives
the chosen order of the two shuffles of the gluon attachments, and the corresponding
$S$ factors.
\begin{minipage}{0.4\textwidth}
		\includegraphics[scale=0.07]{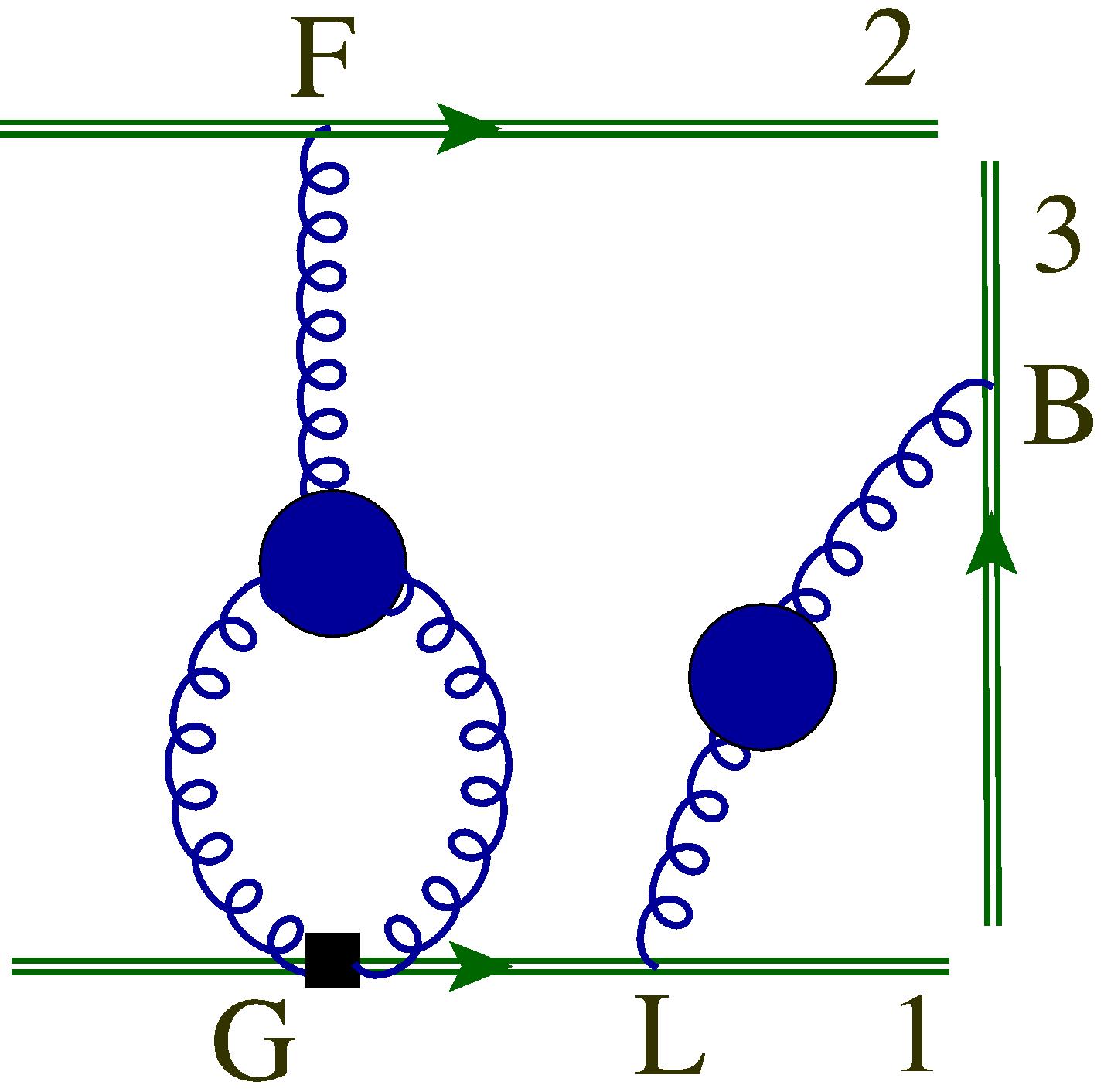}
\end{minipage} 
\begin{minipage}{0.46\textwidth}
	\begin{tabular}{ | c | c | c |}
		\hline
		\textbf{Diagrams} & \textbf{Sequences} & \textbf{S-factors} \\ \hline
		$C_1$ & $\{\lbrace LG\rbrace\}$ & 1 \\ \hline
		$C_2$ & $\{\lbrace GL\rbrace\}$ & 1 \\ \hline
	\end{tabular}	
\end{minipage}

\vspace{0.5cm}
\noindent The $R$, $Y$ and $D$ matrices are given by
\begin{align}
	R=\left(
	\begin{array}{cc}
		\frac{1}{2} & -\frac{1}{2} \\
		-\frac{1}{2} & \frac{1}{2} \\
	\end{array}\right)\,, \qquad Y=\left(
	\begin{array}{cc}
		-1 & 1 \\
		1 & 1 \\
	\end{array}\right)\,,\qquad	\qquad 
	D = \D{1} \, .
	\label{eq:3Web5}
\end{align}
Finally, the exponentiated colour factors are
\begin{align*}
	(YC)_1 =
	f^{abc} f^{bdk} \sck 1 \td 2 \ta 3 
	+ f^{abc} f^{cdk}\skb 1 \td 2 \ta 3 \\
\end{align*}

	\vspace{2mm}
	%%%%%%%%%%%%%%%%%%%%%%%%%%%%%%%%%%%%%%%%%%%%%%%%%%%%%%%%%%%%%%%%%%%%%%%%%%%%%

	\item[{\bf 5}.]  $\textbf{W}_{3}^{(1,1)}(\overline{1},2,1)$
	
This Cweb has two diagrams, one of which is displayed below. The table gives
the chosen order of the two shuffles of the gluon attachments, and the corresponding
$S$ factors.
\begin{minipage}{0.4\textwidth}
		\includegraphics[scale=0.07]{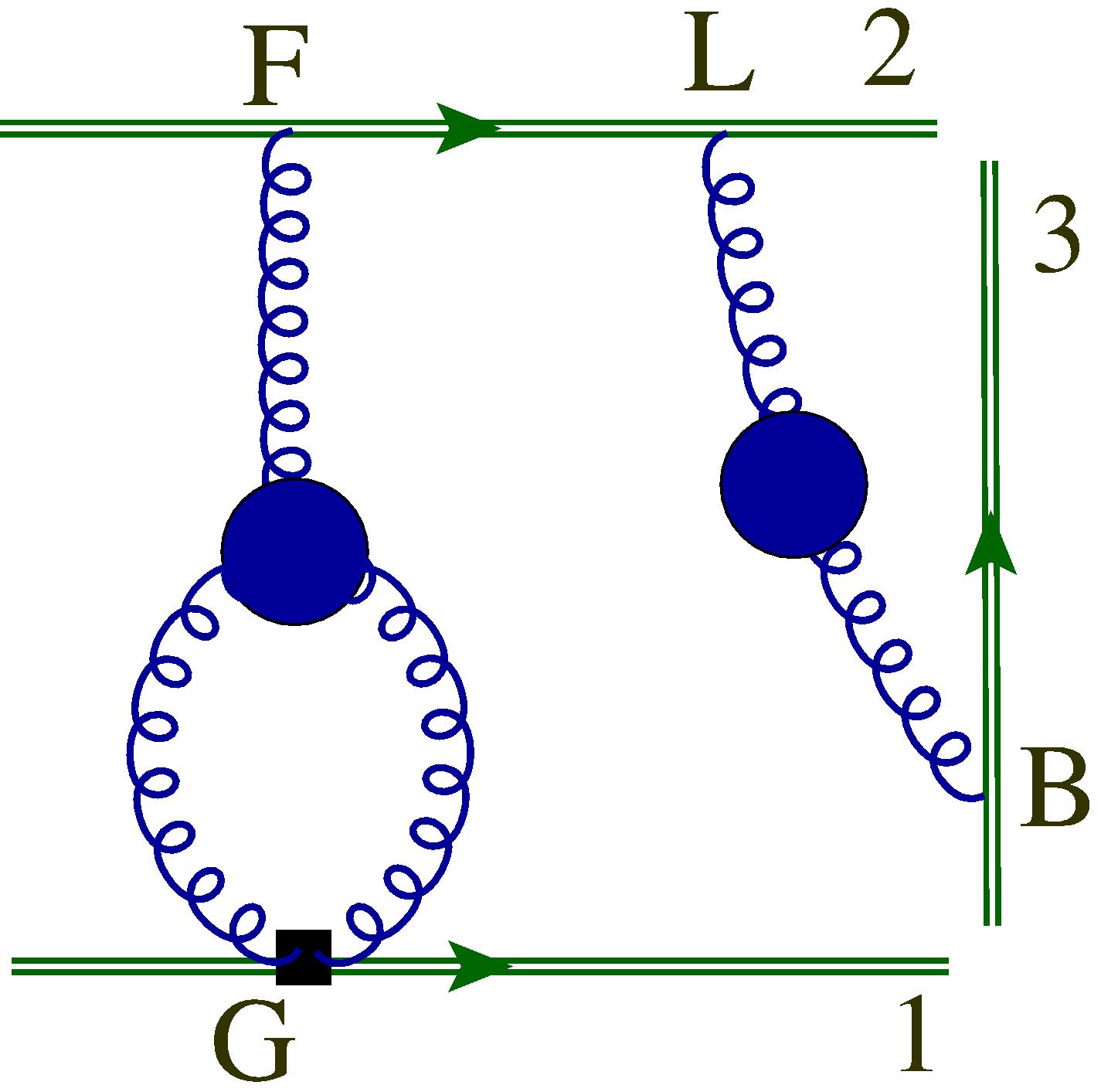}
\end{minipage} 
\begin{minipage}{0.46\textwidth}
	\begin{tabular}{ | c | c | c |}
		\hline
		\textbf{Diagrams} & \textbf{Sequences} & \textbf{S-factors} \\ \hline
		$C_1$ & $\{\lbrace LF\rbrace\}$ & 1 \\ \hline
		$C_2$ & $\{\lbrace FL\rbrace\}$ & 1 \\ \hline
	\end{tabular}	
\end{minipage}

\vspace{0.5cm}
\noindent The $R$, $Y$ and $D$ matrices are given by
\begin{align}
	R=\left(
	\begin{array}{cc}
		\frac{1}{2} & -\frac{1}{2} \\
		-\frac{1}{2} & \frac{1}{2} \\
	\end{array}\right)\,, \qquad Y=\left(
	\begin{array}{cc}
		-1 & 1 \\
		1 & 1 \\
	\end{array}\right)\,,\qquad	\qquad 
	D = \D{1} \, .
	\label{eq:3Web6}
\end{align}
Finally, the exponentiated colour factors are
\begin{align*}
	(YC)_1 =
	f^{abc} f^{adk} \scb 1 \td 2 \tk 3 
\end{align*} 	
	\vspace{2mm}
	%%%%%%%%%%%%%%%%%%%%%%%%%%%%%%%%%%%%%%%%%%%%%%%%%%%%%%%%%%%%%%%%%%%%%%%%%%%%%	
	
\end{itemize}

\subsection{Cwebs with $ S= \{0,0,\cdots, 1_2 \} $ }

The remaining five Cwebs out of ten belong to this category.  Now we will write the form of mixing matrix and its rank for the NE Cwebs contributing at three loops belonging to this class.

\begin{itemize}

	\vspace{2mm}
	
	%%%%%%%%%%%%%%%%%%

	\item[{\bf 1}.]  $\textbf{W}_{3, \, \text{I}}^{(1,1)}(\overline{2},2,1)$
	
	This Cweb has four diagrams, one of which is displayed below. The table gives
	the chosen order of the four shuffles of the gluon attachments, and the corresponding
	$s$-factors.
	\begin{minipage}{0.4\textwidth}
			\includegraphics[scale=0.08]{3Web1}
	\end{minipage} 
	\begin{minipage}{0.46\textwidth}
		\begin{tabular}{ | c | c | c |}
			\hline
			\textbf{Diagrams} & \textbf{Sequences} & \textbf{S-factors} \\ \hline
			$C_1$ & $\{\lbrace GL \rbrace, \lbrace AF\rbrace\}$ & 0\\ \hline
			$C_2$ & $\{\lbrace LG \rbrace, \lbrace FA\rbrace\}$ & 0\\ \hline
			$C_3$ & $\{\lbrace GL\rbrace, \lbrace FA\rbrace\}$ & 1 \\ \hline
			$C_4$ & $\{\lbrace LG \rbrace, \lbrace AF\rbrace\}$ & 1\\ \hline
		\end{tabular}	
	\end{minipage}
	
	\noindent The mixing matrix and its rank are given by
	\begin{align}
		R=R(0_2,1_2)    \,\qquad	\qquad 
		r(R(0_2,1_2)) = 4-1 = 3  \, .
	\end{align}

\begin{align}
	Y = \left(
	\begin{array}{cccc}
		-1 & 0 & 0 & 1 \\
		-1 & 0 & 1 & 0 \\
		-1 & 1 & 0 & 0 \\
		0 & 0 & 1 & 1 \\
	\end{array}
	\right)
\end{align}
The exponentiated colour factors are
	\begin{align*}
	(YC)_1 &=
	- i f^{acn}\sba 1 \tc 1 \tb 2 \tn 3  
	- i f^{ack}\sbk 1 \tb 2 \tc 3 \ta 3 
	- i f^{bck} \ska 1 \tb 2 \tc 3 \ta 3 
	\\
	(YC)_2 &=
	-i f^{ack} \sbk 1 \tb 2 \ta 3 \tc 3 
	-i f^{bck} \ska 1 \tb 2 \ta 3 \tc 3 \\
	(YC)_3 &=
	-i f^{acn} \sba 1 \tc 1 \tb 2 \tn 3 
\end{align*}

	\vspace{2mm}
	
	%%%%%%%%%%%%%%%%%%

	\item[{\bf 2}.]  $\textbf{W}_{3, \, \text{II}}^{(1,1)}(\overline{1},2,2)$
	
	This Cweb has four diagrams, one of which is displayed below. The table gives
	the chosen order of the four shuffles of the gluon attachments, and the corresponding
	$s$-factors.
	\begin{minipage}{0.4\textwidth}
			\includegraphics[scale=0.08]{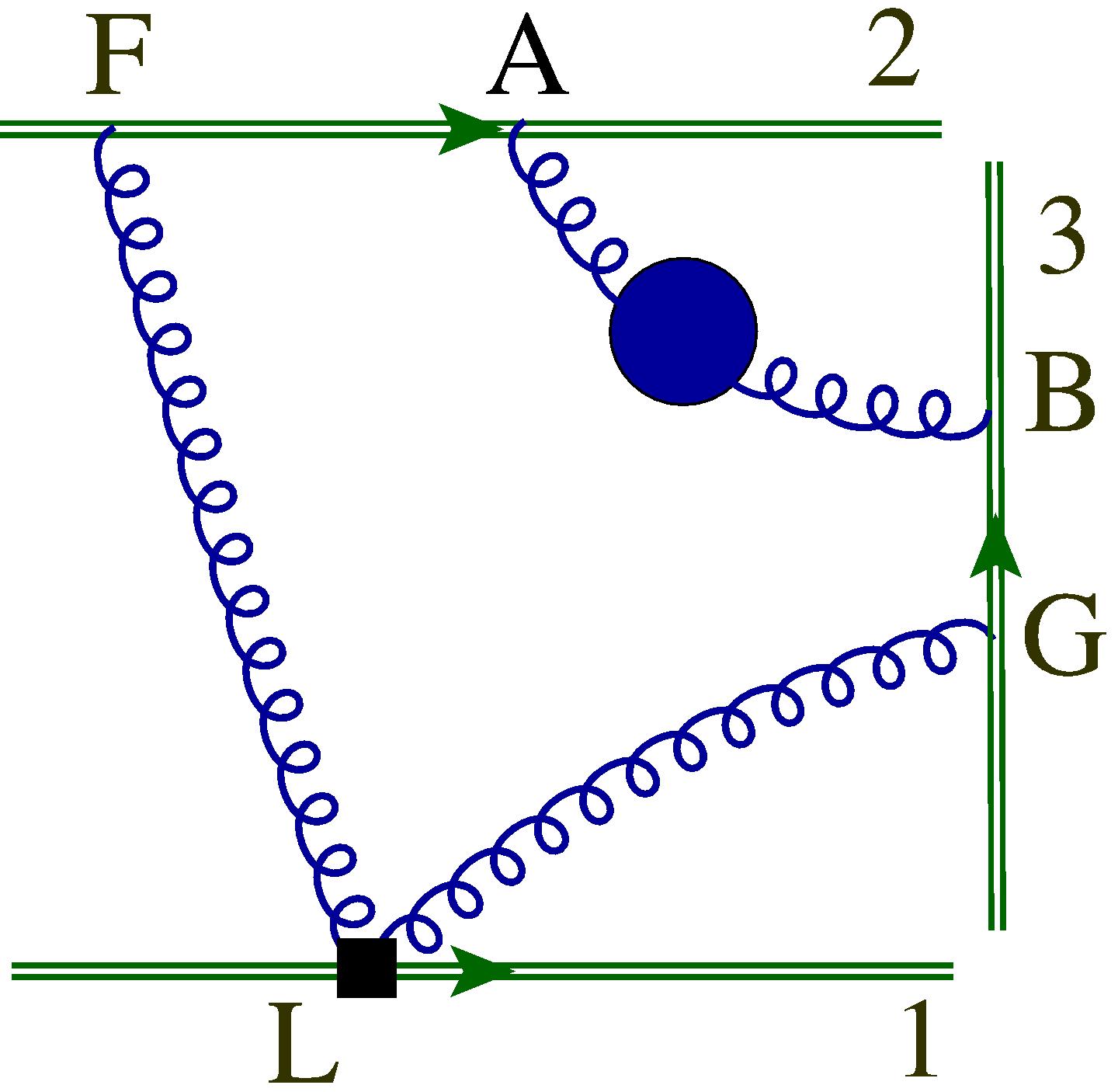}
	\end{minipage} 
	\begin{minipage}{0.46\textwidth}
		\begin{tabular}{ | c | c | c |}
			\hline
			\textbf{Diagrams} & \textbf{Sequences} & \textbf{S-factors} \\ \hline
			$C_1$ & $\{\lbrace BG \rbrace, \lbrace FA\rbrace\}$ & 0\\ \hline
			$C_2$ & $\{\lbrace GB \rbrace, \lbrace AF\rbrace\}$ & 0\\ \hline
			$C_3$ & $\{\lbrace BG\rbrace, \lbrace AF\rbrace\}$ & 1 \\ \hline
			$C_4$ & $\{\lbrace GB \rbrace, \lbrace FA\rbrace\}$ & 1\\ \hline
		\end{tabular}	
	\end{minipage}
	
	\vspace{0.5cm}
	\noindent The mixing matrix and its rank are given by
	\begin{align}
		R=R(0_2,1_2)    \,\qquad	\qquad 
		r(R(0_2,1_2)) = 4-1 = 3  \, .
	\end{align}

	\begin{align}
		Y = \left(
		\begin{array}{cccc}
			-1 & 0 & 0 & 1 \\
			-1 & 0 & 1 & 0 \\
			-1 & 1 & 0 & 0 \\
			0 & 0 & 1 & 1 \\
		\end{array}
		\right)
	\end{align}

	Finally, the exponentiated colour factors are
\begin{align*}
	(YC)_1 &=
	-i f^{bcn} \sab 1 \tc 2 \ta 2 \tn 3 
	- i f^{ack}\sab 1 \tk 2 \tb 3 \tc 3 
	\\
	(YC)_2 &=
	- i f^{ack}\sab 1 \tk 2 \tb 3 \tc 3 
	\\
	(YC)_3 &=
	-i f^{bcn} \sab 1 \ta 2 \tc 2 \tn 3 
\end{align*} 

	\vspace{2mm}
	
	%%%%%%%%%%%%%%%%%%
	%%%%%%%%%%%%%%%%%%

	\item[{\bf 3}.]  $\textbf{W}_{3}^{(1,1)}(\overline{1},3,1)$
	
	This Cweb has three diagrams, one of which is displayed below. The table gives
	the chosen order of the three shuffles of the gluon attachments, and the corresponding
	$s$-factors.
	\begin{minipage}{0.4\textwidth}
			\includegraphics[scale=0.08]{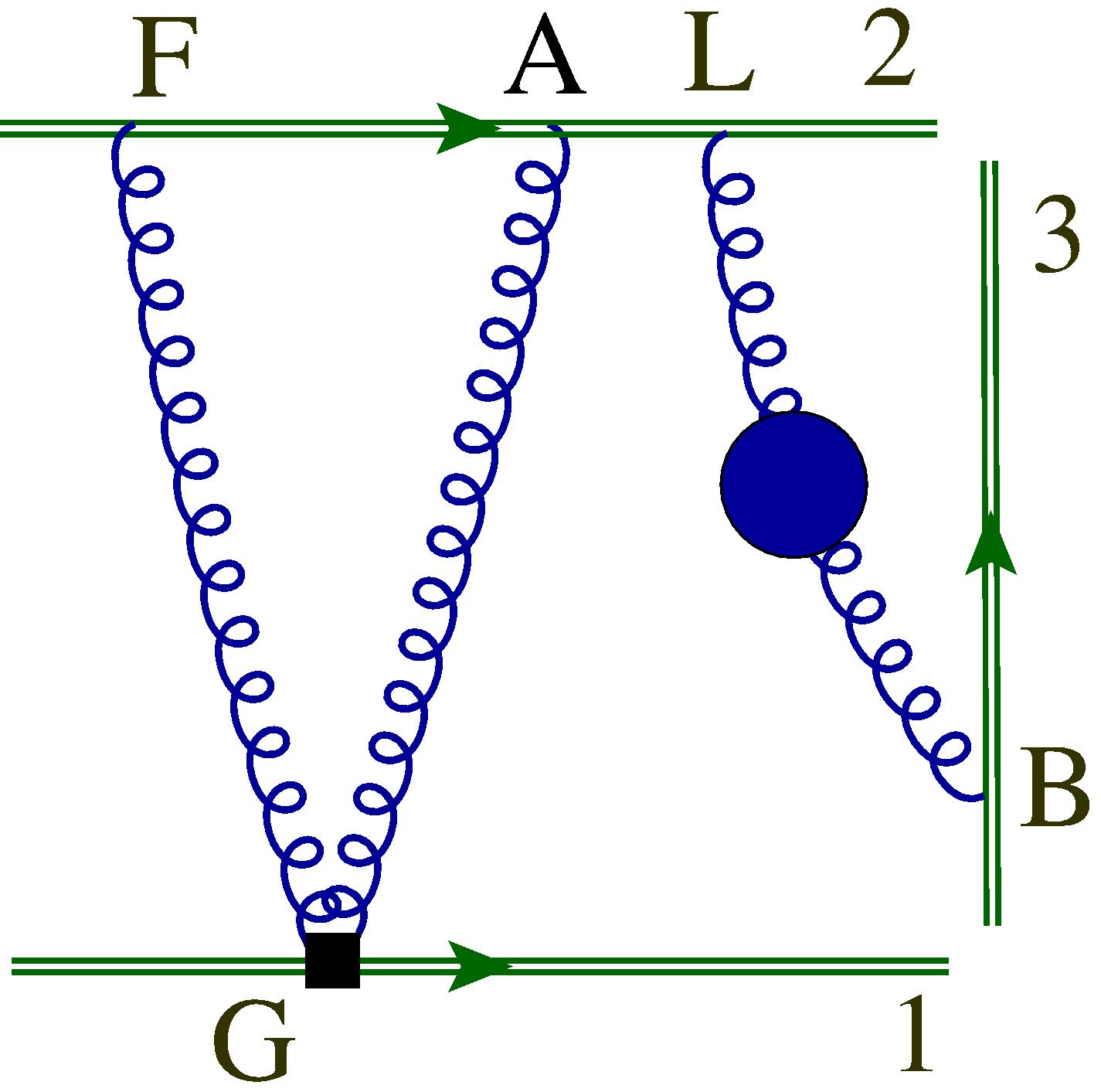}
	\end{minipage} 
	\begin{minipage}{0.46\textwidth}
		\begin{tabular}{ | c | c | c |}
			\hline
			\textbf{Diagrams} & \textbf{Sequences} & \textbf{S-factors} \\ \hline
			$C_1$ & $\{\lbrace FLA\rbrace\}$ & 0 \\ \hline
			$C_2$ & $\{\lbrace LFA\rbrace\}$ & 1 \\ \hline
			$C_3$ & $\{\lbrace FAL \rbrace\}$ & 1\\ \hline
		\end{tabular}	
	\end{minipage}
	
	\vspace{0.5cm}
	\noindent The mixing matrix and its rank are given by
	\begin{align}
		R=R(0_1,1_2)    \,\qquad	\qquad 
		r(R(0_1,1_2)) = 3-1 = 2  \, .
	\end{align}
	\begin{align}
		Y =  \left(
		\begin{array}{ccc}
			-1 & 0 & 1 \\
			-1 & 1 & 0 \\
			0 & 1 & 1 \\
		\end{array}
		\right)
	\end{align}
	Finally, the exponentiated colour factors are
\begin{align*}
	(YC)_1 &=
	- i f^{ack} \sba 1 \tc 2 \tb 3 \tk 3 
	- i f^{bcn} \sba 1 \tc 2 \tn 3 \ta 3  \\
	(YC)_2 &=
	- i f^{ack} \sba 1 \tc 2 \tb 3 \tk 3 
\end{align*}

	\vspace{2mm}
	
	%%%%%%%%%%%%%%%%%%

	%%%%%%%%%%%%%%%%%%

	%%%%%%%%%%%%%%%%%%%%%%%%%%%%%%%%%%%%%%%%%%%%%%%%%%%%%%%%%%%%%%%%%%%%%%%%%%%%%

	\item[{\bf 4}.]  $\textbf{W}_{2}^{(1,1)}(\overline{2},3)$
	
	This Cweb has six diagrams, one of which is displayed below. The table gives
	the chosen order of the six shuffles of the gluon attachments, and the corresponding
	$s$-factors.
	\begin{minipage}{0.4\textwidth}
			\includegraphics[scale=0.08]{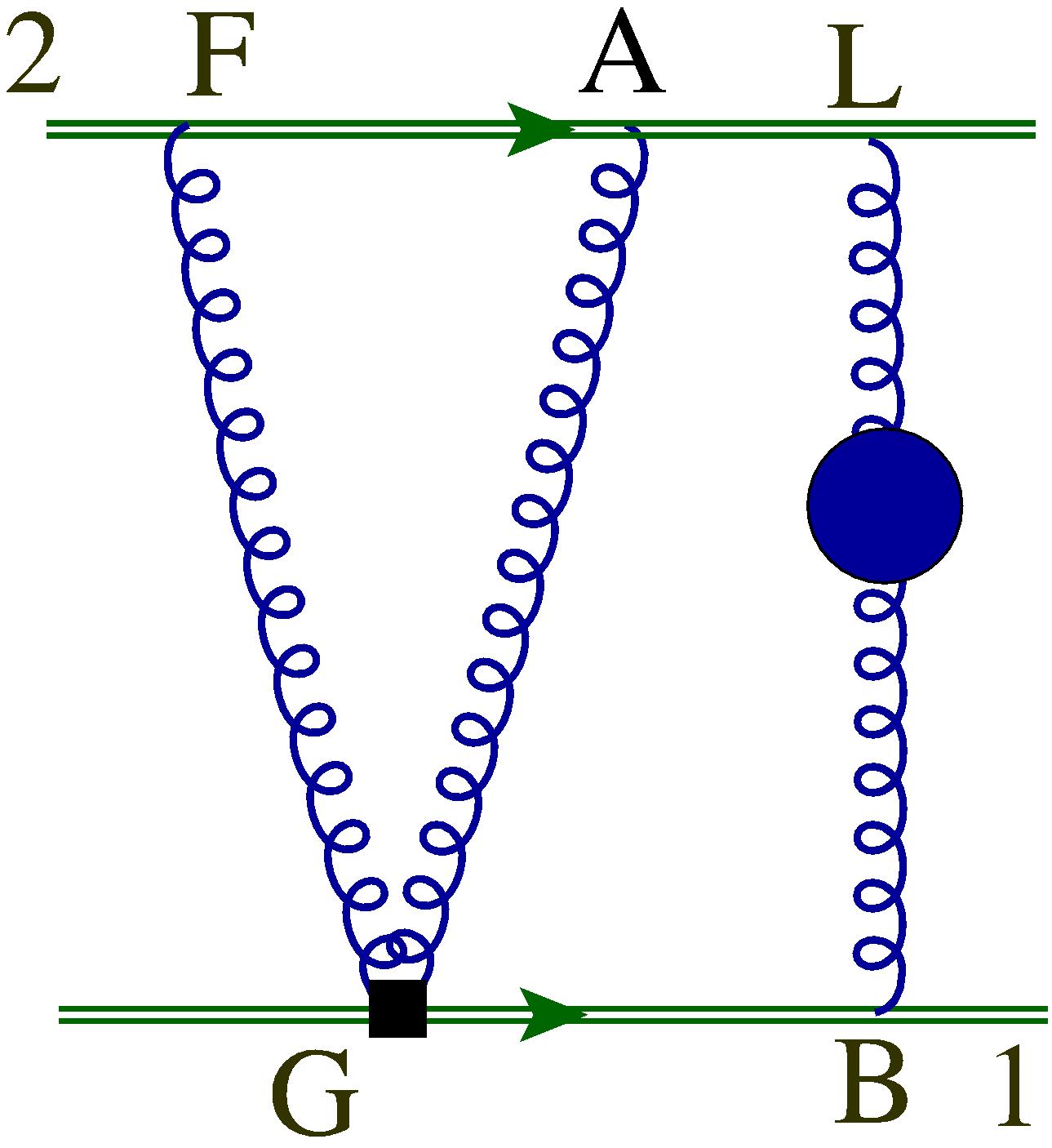}
	\end{minipage} 
	\begin{minipage}{0.46\textwidth}
		\begin{tabular}{ | c | c | c |}
			\hline
			\textbf{Diagrams} & \textbf{Sequences} & \textbf{S-factors} \\ \hline
			$C_1$ & $\{\lbrace BG\rbrace, \lbrace FLA\rbrace\}$ & 0 \\ \hline
			$C_2$ & $\{\lbrace BG\rbrace, \lbrace FAL\rbrace\}$ & 0 \\ \hline
			$C_3$ & $\{\lbrace GB\rbrace, \lbrace LFA\rbrace\}$ & 0 \\ \hline
			$C_4$ & $\{\lbrace GB\rbrace, \lbrace FLA\rbrace\}$ & 0 \\ \hline
			$C_5$ & $\{\lbrace BG\rbrace, \lbrace LAF\rbrace\}$ & 1 \\ \hline
			$C_6$ & $\{\lbrace GB\rbrace, \lbrace FAL\rbrace\}$ & 1 \\ \hline
		\end{tabular}	
	\end{minipage}
	
	\vspace{0.5cm}
	\noindent The mixing matrix and its rank are given by
	\begin{align}
		R=R(0_4,1_2)    \,\qquad	\qquad 
		r(R(0_4,1_2)) = 6-1 = 5  \, .
	\end{align} 

\begin{align}
	Y = \left(
	\begin{array}{cccccc}
		-1 & 0 & 0 & 0 & 0 & 1 \\
		-1 & 0 & 0 & 0 & 1 & 0 \\
		-1 & 0 & 0 & 1 & 0 & 0 \\
		-1 & 0 & 1 & 0 & 0 & 0 \\
		-1 & 1 & 0 & 0 & 0 & 0 \\
		0 & 0 & 0 & 0 & 1 & 1 \\
	\end{array}
	\right)
\end{align}

		Finally, the exponentiated colour factors are
	\begin{align*}
		(YC)_1 &=
		- i f^{bck}\tc 1 \sab 1 \ta 2 \tk 2  
		- i f^{ack} \tc 1 \sab 1 \tk 2 \tb 2 
		-i f^{bcm} \sam 1 \ta 2 \tb 2 \tc 2  
		-i f^{acm} \smb 1 \ta 2 \tb 2 \tc 2 \\
		(YC)_2 &=
		-i f^{bck} \tc 1 \sab 1 \ta 2 \tk 2 
		- i f^{bcm}\sam 1 \ta 2 \tb 2 \tc 2
		-i f^{acm} \smb 1 \ta 2 \tb 2 \tc 2 \\
		(YC)_3 &=
		-i f^{bcm} \sam 1 \ta 2 \tb 2 \tc 2 
		-i f^{acm} \smb 1 \ta 2 \tb 2 \tc 2 \\
		(YC)_4 &=
		-if^{bck} \sab 1 \tc 1 \ta 2 \tk 2 
		-i f^{ack} \sab 1 \tc 1 \tk 2 \tb 2 \\
		(YC)_5 &=
		-i f^{bck} \sab 1 \tc 1 \ta 2 \tk 2 
	\end{align*} 
	
	\vspace{2mm}
	%%%%%%%%%%%%%%%%%%%%%%%%%%%%%%%%%%%%%%%%%%%%%%%
	
	\item[{\bf 5}.]  $\textbf{W}_{2}^{(1,1)}(\overline{2},2)$
	
	This Cweb has four diagrams, one of which is displayed below. The table gives
	the chosen order of the four shuffles of the gluon attachments, and the corresponding
	$s$-factors.
	\begin{minipage}{0.4\textwidth}
			\includegraphics[scale=0.08]{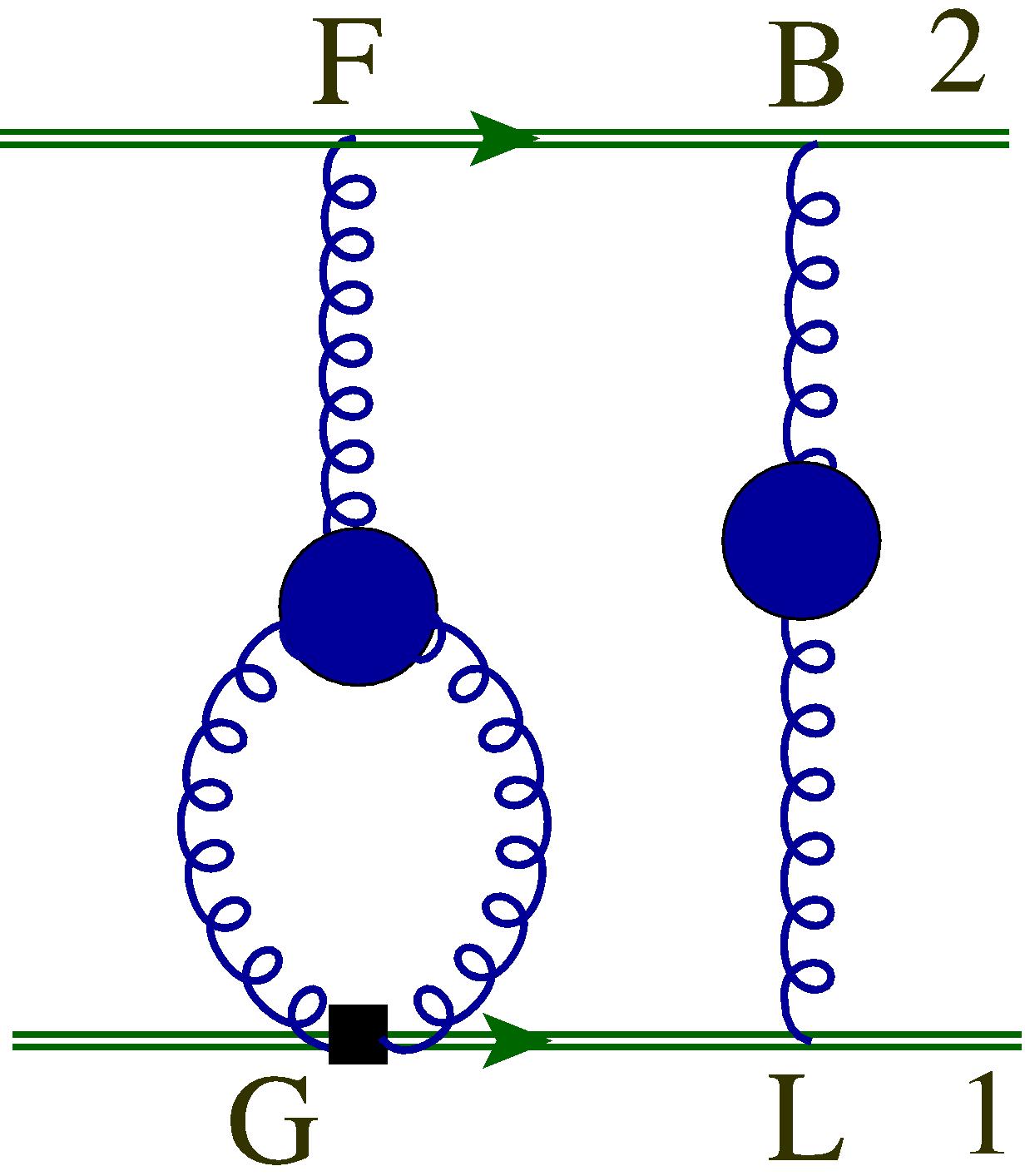}
	\end{minipage} 
	\begin{minipage}{0.42\textwidth}
		\begin{tabular}{ | c | c | c |}
			\hline
			\textbf{Diagrams} & \textbf{Sequences} & \textbf{S-factors} \\ \hline
			$C_1$ & $\{\lbrace LG\rbrace, \lbrace FB\rbrace\}$ & 0 \\ \hline
			$C_2$ & $\{\lbrace GL\rbrace, \lbrace BF\rbrace\}$ & 0 \\ \hline
			$C_3$ & $\{\lbrace LG\rbrace, \lbrace BF\rbrace\}$ & 1 \\ \hline
			$C_4$ & $\{\lbrace GL\rbrace, \lbrace FB\rbrace\}$ & 1 \\ \hline
		\end{tabular}	
	\end{minipage}
	
	\vspace{0.5cm}
	\noindent The mixing matrix and its rank are given by
	\begin{align}
		R=R(0_2,1_2)    \,\qquad	\qquad 
		r(R(0_2,1_2)) = 4-1 = 3  \, .
	\end{align}

\begin{align}
	Y = \left(
	\begin{array}{cccc}
		-1 & 0 & 0 & 1 \\
		-1 & 0 & 1 & 0 \\
		-1 & 1 & 0 & 0 \\
		0 & 0 & 1 & 1 \\
	\end{array}
	\right)
\end{align}
	Finally, the exponentiated colour factors are
\begin{align*}
	(YC)_1 &=
	f^{abc} f^{adk} \scb 1 \td 1 \tk 2 
	+ f^{abc} f^{bdk} \sck 1 \td 2 \ta 2 
	+ f^{abc} f^{cdk} \skb 1 \td 2 \ta 2 \\
	(YC)_2 &=
	f^{abc} f^{bdk} \sck 1 \ta 2 \td 2 f^{abc} f^{bdk}
	+ f^{abc} f^{cdk} \skb 1 \ta 2 \td 2 \\
	(YC)_3 &= f^{abc} f^{adk}
	\scb 1 \td 1 \tk 2 \\
\end{align*} 

\end{itemize}

\section{Completely connected NE Webs}\label{sec:ConnectedWebs}

Out of the nineteen NE Cwebs that appear at three-loops, nine of them are fully connected having only one diagram, and thus have a trivial mixing matrix $ R=1 $. We have listed these one-diagram Cwebs here.  

\begin{figure}[h]
	\centering
	\subfloat[][$\textbf{W}_{2}^{(0,1)}(\overline{2},1)$]{\includegraphics[scale=0.07]{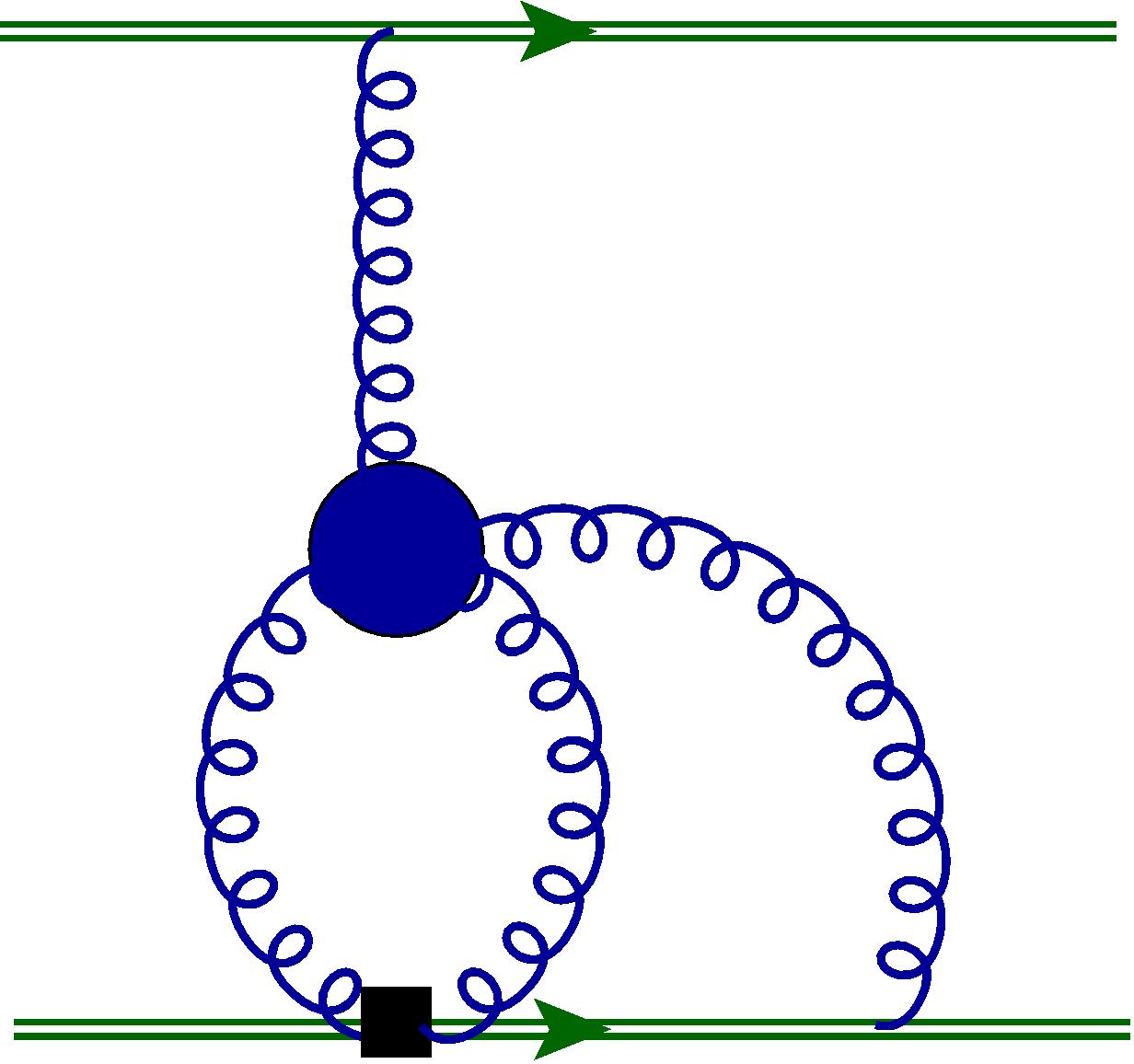} }
	\qquad 
	\subfloat[][ $\textbf{W}_{2}^{(0,1)}(\overline{1},2)$]{\includegraphics[scale=0.07]{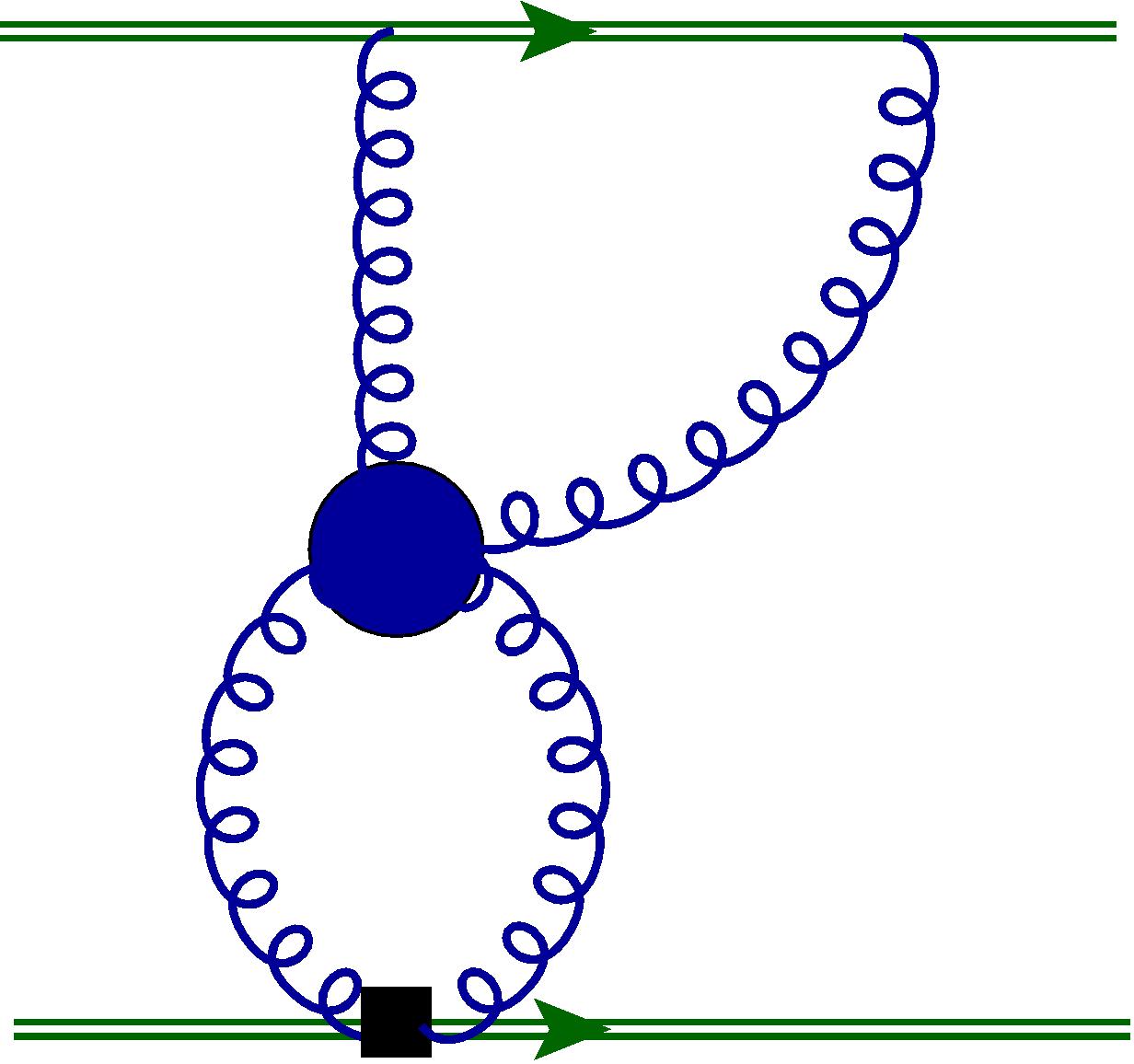} }
	\qquad \quad
	\subfloat[][$\textbf{W}_{2,\text{II}}^{(0,0,1)}(\overline{2},2)$]{\includegraphics[scale=0.067]{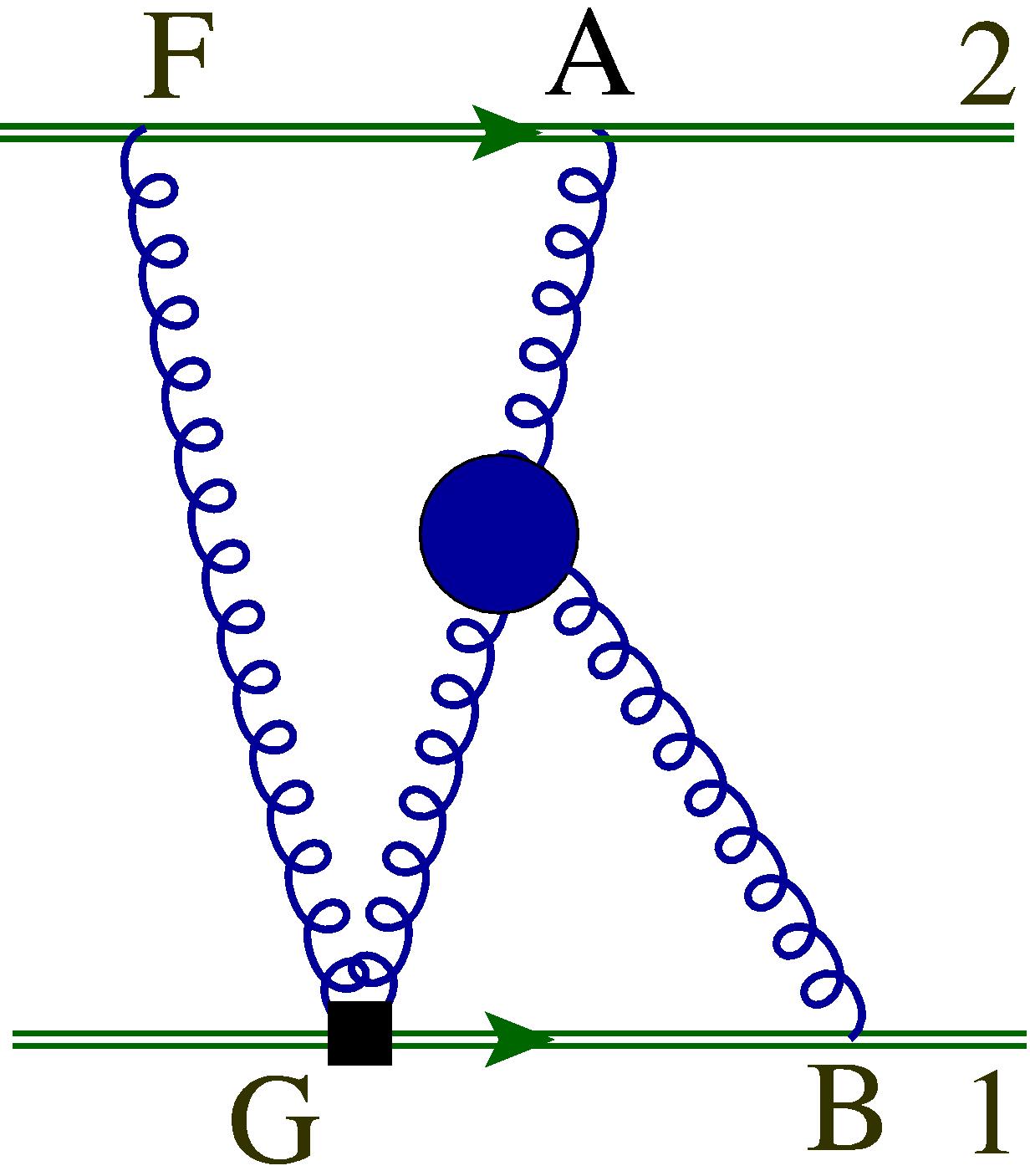} }
	\qquad 
	\subfloat[][$\textbf{W}_{2}^{(0,0,1)}(\overline{1},3)$]{\includegraphics[scale=0.067]{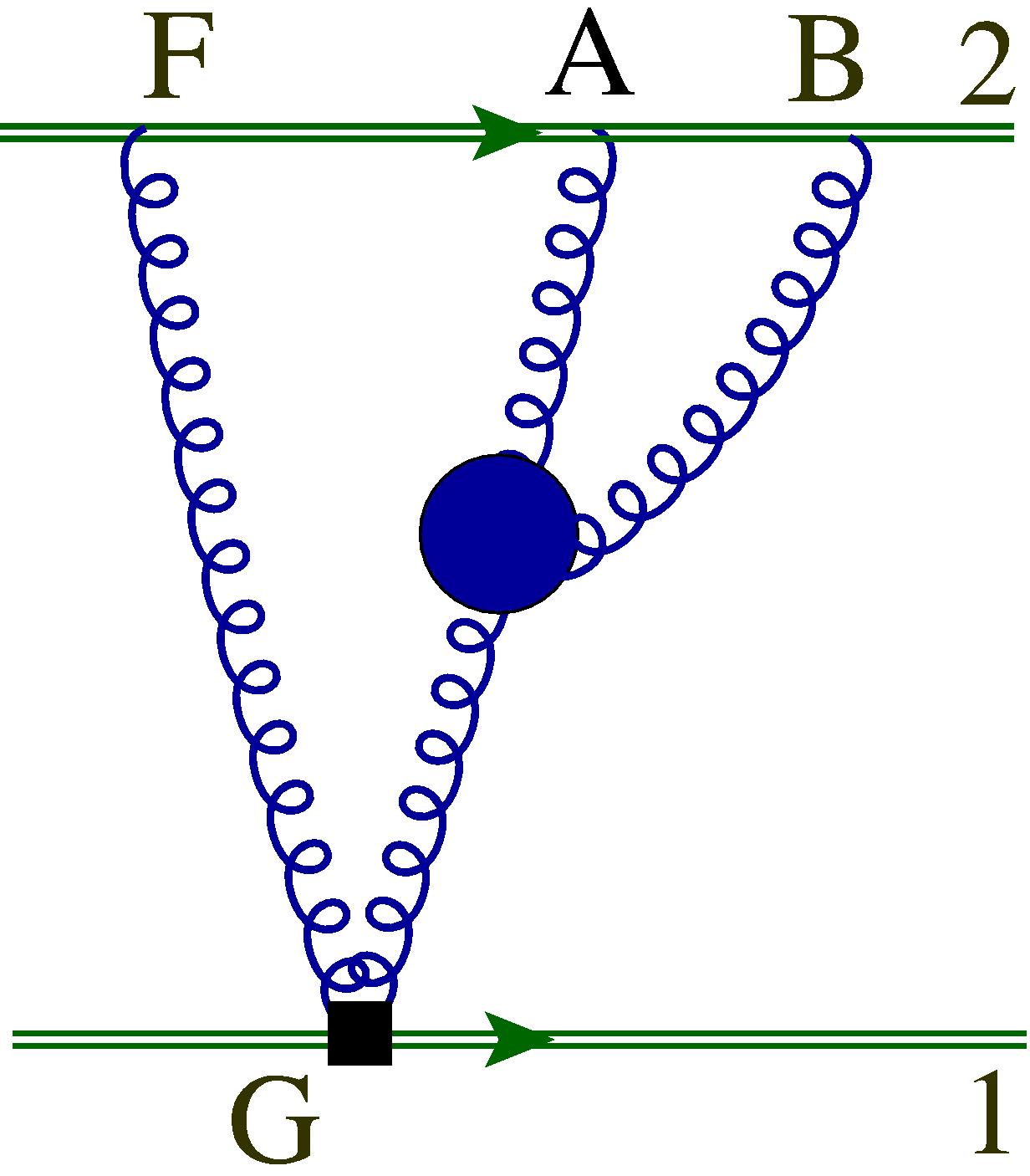} }
	\qquad 
	\subfloat[][$\textbf{W}_{3}^{(0,0,1)}(\overline{1},2,1)$]{\includegraphics[scale=0.07]{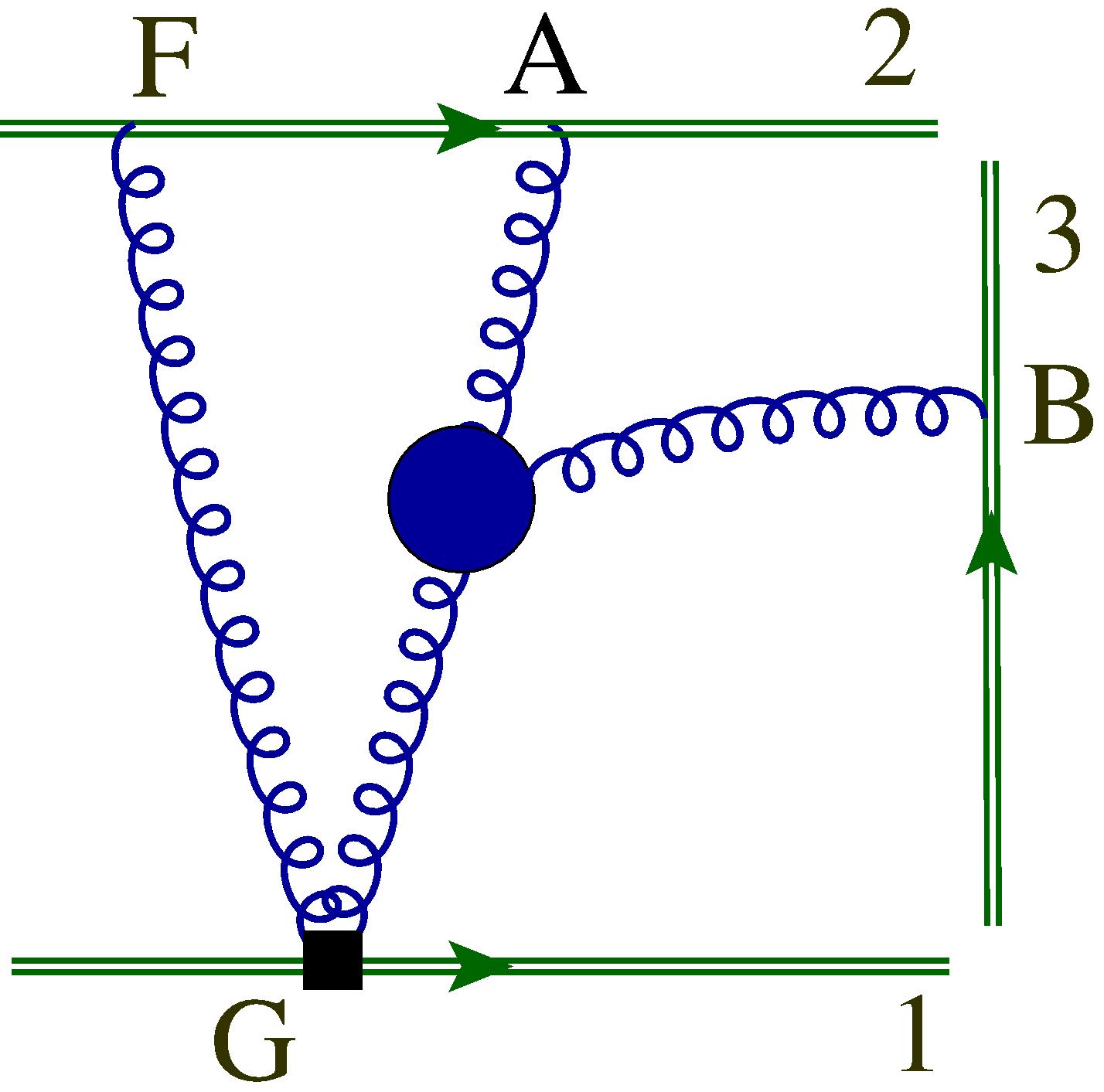} }
	\qquad 
	\subfloat[][$\textbf{W}_{3}^{(0,0,1)}(\overline{1},1,2)$]{\includegraphics[scale=0.07]{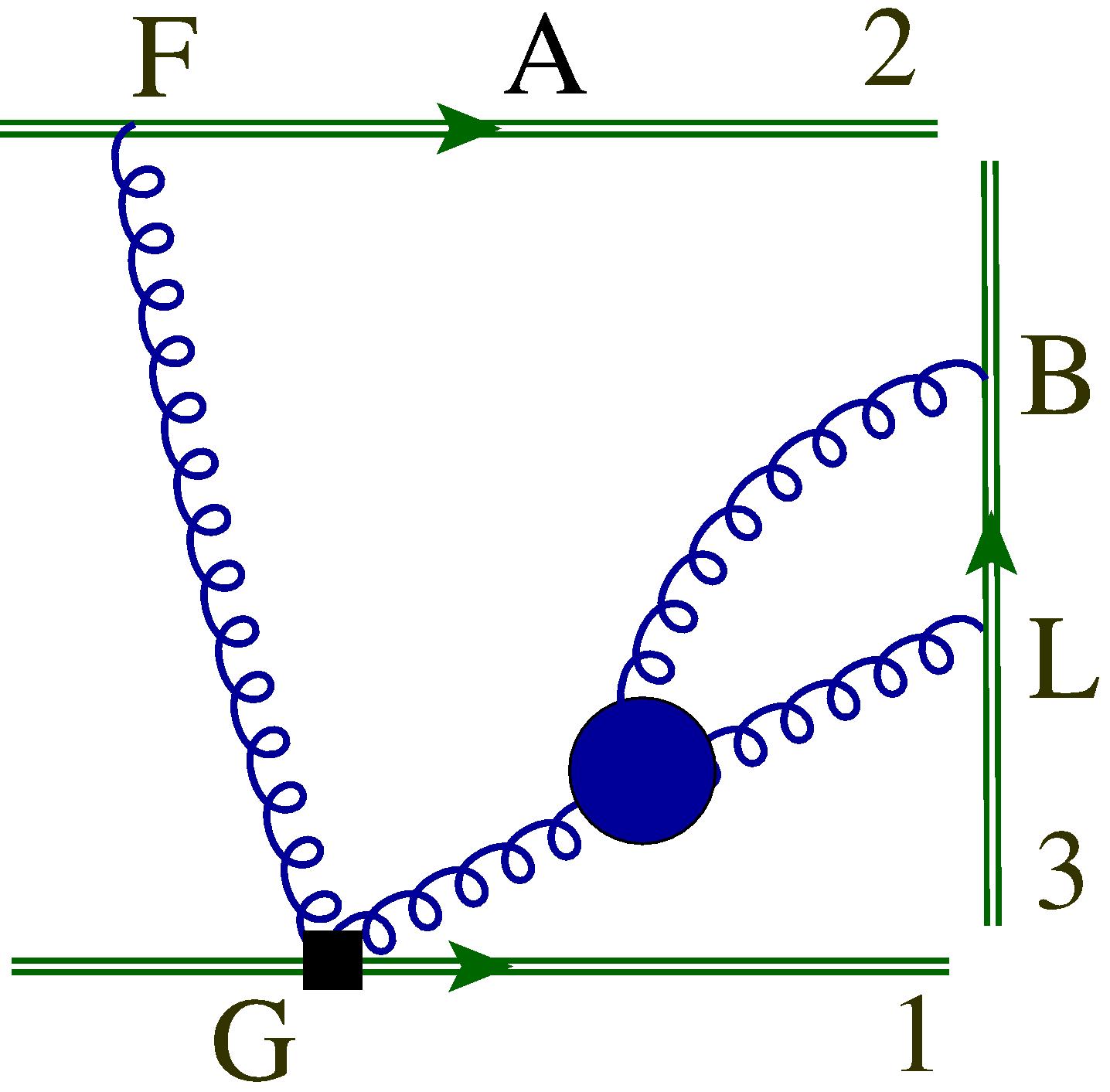} }
	\qquad 
	\subfloat[][$\textbf{W}_{3}^{(0,0,1)}(\overline{2},1,1)$]{\includegraphics[scale=0.07]{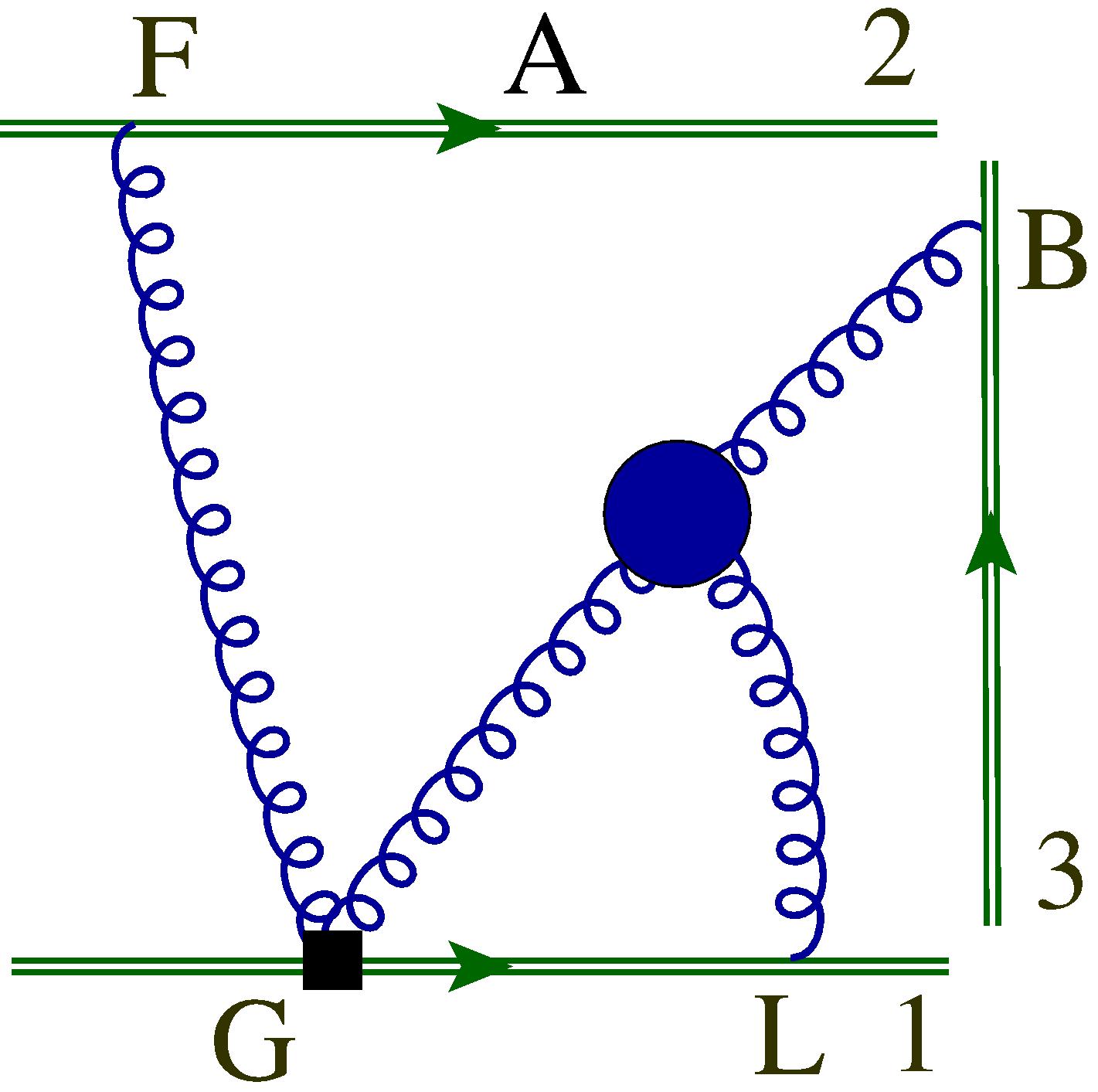} }   
	\qquad 
	\subfloat[][$\textbf{W}_{3}^{(0,1)}(\overline{1},1,1)$]{\includegraphics[scale=0.07]{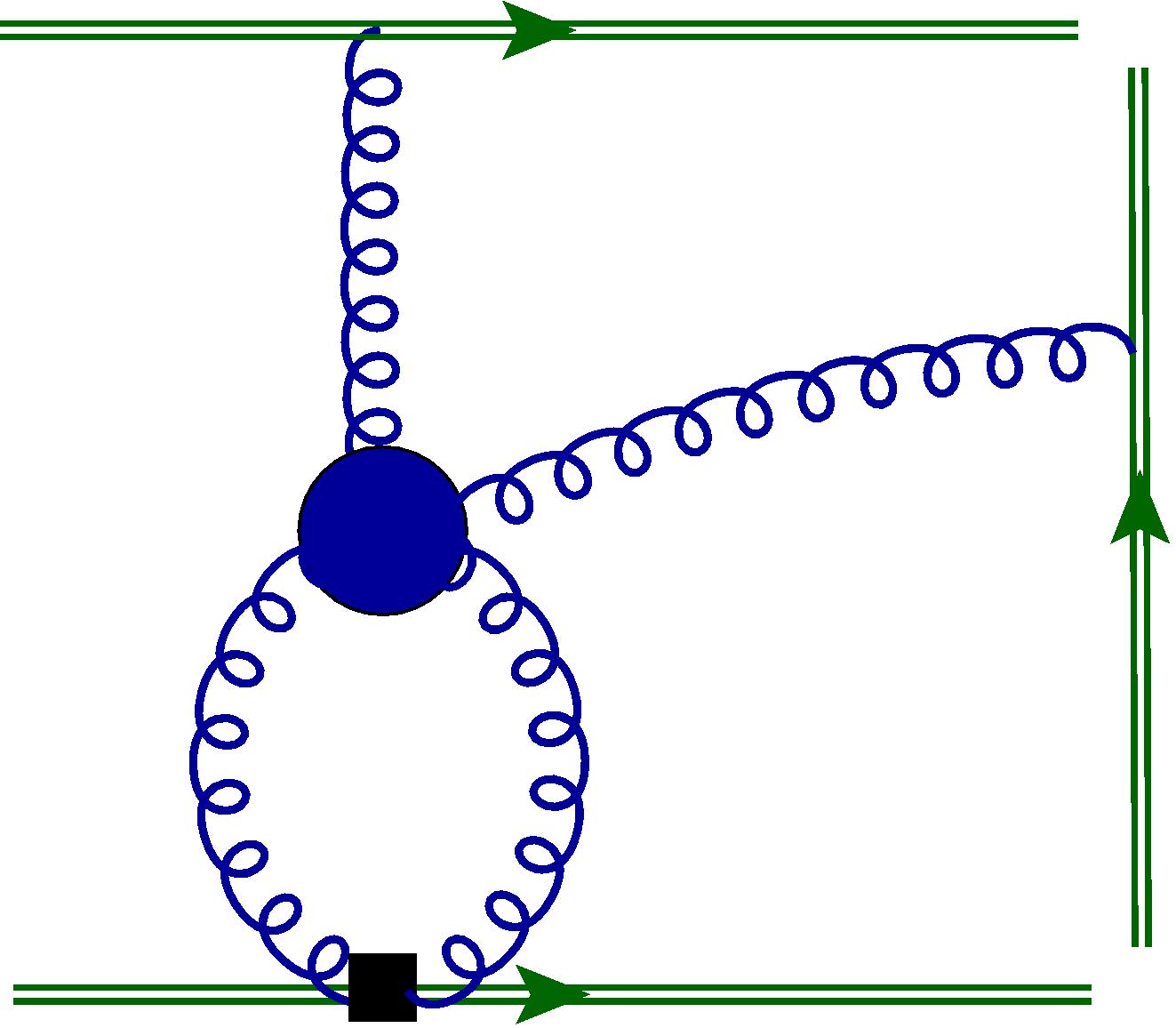} }
	\qquad 
	\subfloat[][$\textbf{W}_{4}^{(0,0,1)}(\overline{1},1,1,1)$]{\includegraphics[scale=0.07]{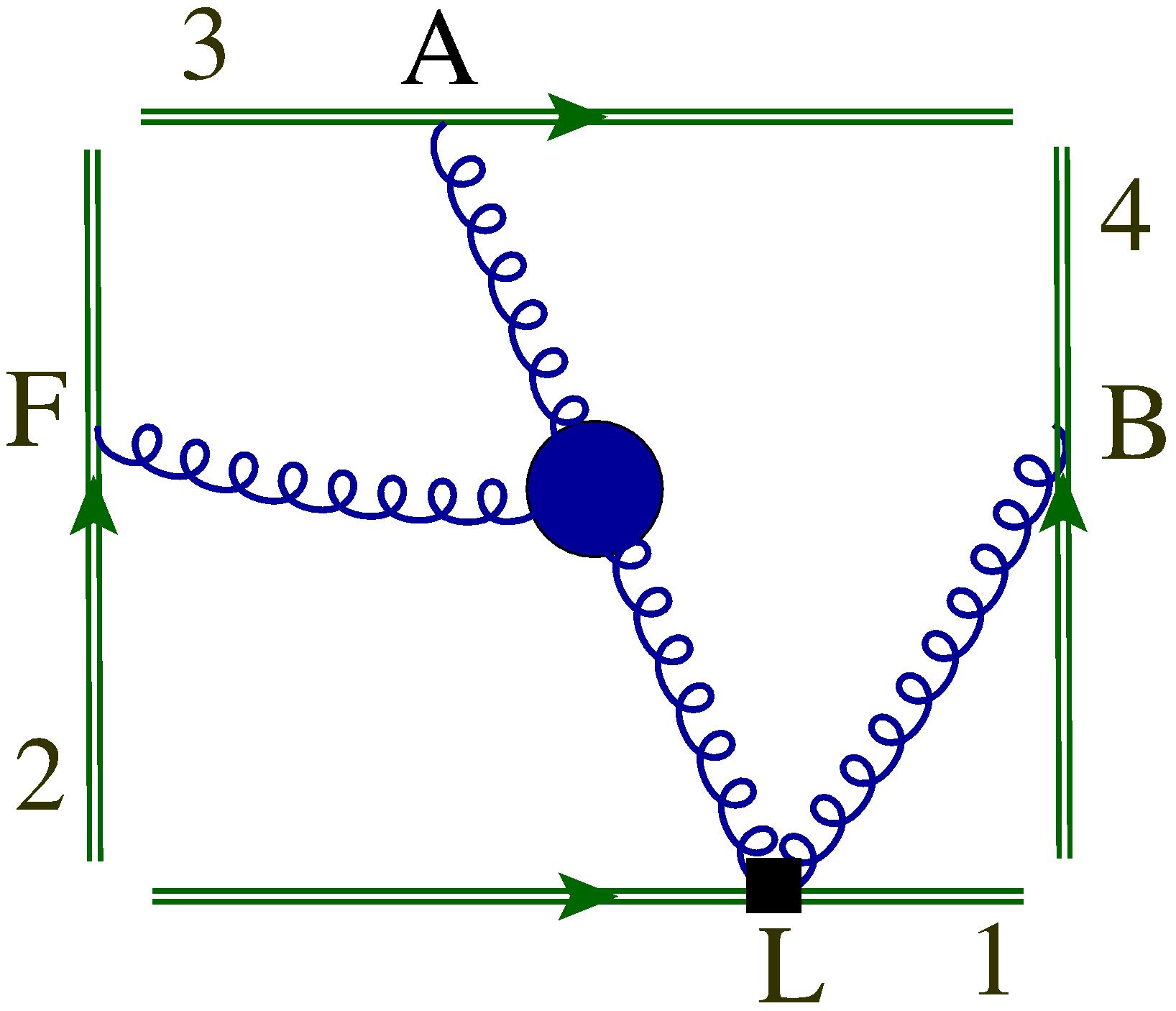} }       
	\caption{NE Cwebs at 3-loop with $R=1$}
	%	\label{fig:Lows}
\end{figure}

%%%%%%%%%%%%%%%%%%%%%%%%%%%%%%%%%%%%%%%%%%%%%%%%%%%%%%%%%%%%%%%%%%%%%
\bibliographystyle{bibstyle}
\bibliography{NEwebs}
\end{document}